\documentclass[lettersize,journal]{IEEEtran}

\usepackage{cite}
\usepackage{easybmat}
\usepackage{stfloats}
\usepackage{amsmath,amssymb,amsfonts}
\usepackage{graphicx}
\usepackage{textcomp}
\usepackage{footmisc}
\usepackage{algorithm}
\usepackage{epstopdf}
\usepackage{balance}
\usepackage{caption}
\usepackage{subcaption}
\usepackage{algorithmic}
\usepackage[dvipsnames]{xcolor}
\usepackage[letterpaper, left=0.7in, right=0.7in, bottom=1in, top=0.7in]{geometry}
\usepackage{booktabs}
\usepackage[printonlyused,withpage]{acronym}
\usepackage{empheq}

\newtheorem{thm}{Theorem}
\newtheorem{prop}{Proposition}
\newtheorem{lem}{Lemma}
\newtheorem{defi}{Definition}

\newtheorem{cor}{Corollary}
\newtheorem{rmk}{Remark}

\newenvironment{proof}{ \paragraph*{Proof}}{\hfill$\square$}
\renewcommand{\baselinestretch}{.93} 

\newcommand{\mrm}[1]{\ensuremath{\text{#1}}}
\renewcommand{\vec}{\ensuremath{\mrm{vec}}}
\newcommand{\diag}{\ensuremath{\mrm{diag}}}

\newcommand{\bs}[1]{\ensuremath{\boldsymbol{#1}}}
\newcommand{\comment}[1]{}

\acrodef{MiLAC}{microwave linear analog computer}
\acrodef{RF}{radio-frequency}
\acrodef{MISO}{multiple-input single-output}
\acrodef{MIMO}{multiple-input multiple-output}
\acrodef{OFDM}{orthogonal frequency-division multiplexing}
\acrodef{IID}{independent and identically distributed}
\acrodef{BS}{base station}
\acrodef{SNR}{signal-to-noise ratio}
\acrodef{AWGN}{additive white Gaussian noise}
\acrodef{MRT}{maximum ratio transmission}
\acrodef{MF}{matched filter}
\acrodef{DoFs}{degrees of freedom}
\acrodef{RHS}{right-hand side}
\acrodef{LHS}{left-hand side}
\acrodef{BB}{baseband}
\acrodef{LoS}{line-of-sight}
\acrodef{mmWave}{millimeter-wave}
\acrodef{ADC}{analog-to-digital converter}
\acrodef{DAC}{digital-to-analog converter}
\acrodef{SVD}{singular-value decomposition}
\begin{document}

\title{Wideband Microwave Linear Analog Computer (MiLAC) for Frequency-Flat and Frequency-Selective Processing}
\author{Juan Vidal Alegr\'{i}a, Matteo Nerini, Bruno Clerckx
\thanks{This work has been partially supported by the Swedish Foundation of Strategic Research (CHI19-0001) and by UKRI grant EP/Y004086/1, EP/X040569/1, EP/Y037197/1, EP/X04047X/1, EP/Y037243/1.\par J. Vidal Alegr\'{i}a is with the Department of Electrical and Information Technology, Lund University, SE-22100 Lund, Sweden (email: juan.vidal\_alegria@eit.lth.se).\par Matteo Nerini and Bruno Clerckx are with the Department of Electrical and Electronic Engineering, Imperial College London, SW7 2AZ London, U.K. (email: m.nerini20@imperial.ac.uk;  b.clerckx@imperial.ac.uk).} }

\maketitle

\begin{abstract}
Microwave linear analog computer (MiLAC) has been proposed as an analog processing architecture to perform efficient computation in the microwave domain. Consisting of reconfigurable impedance networks, a MiLAC can extract many of the gains of large multi-antenna systems without costly digital processing, e.g., achieving capacity in narrowband multiple-input single-output (MISO). However, the behavior of wideband MiLAC, required to support high data-rate transmissions, is widely unknown. The impedance components of MiLACs have non-trivial frequency-dependent response, requiring specific implementations to characterize it. In this work, we consider MiLAC implementations based on LC resonators with tunable capacitance, which allow synthesizing near-arbitrary passive lossless impedance matrices at individual frequencies using inductors and varactors. We characterize the mapping between optimal MISO precoding at a reference frequency, and the vector of capacitances configuring the MiLAC. We derive the degrees of freedom (DoFs) remaining after fixing such optimum precoding at the reference frequency, and show that these DoFs can be exploited to tune the frequency response of the MiLAC. We propose two techniques to exploit these DoFs: a closed-form approach for frequency-flat MiLAC processing, and a numerical approach for multi-frequency MiLAC processing, optimizing processing at frequencies other than the reference one. The frequency-flat approach shows satisfactory performance in a simulated frequency-flat millimeter-wave scenario, approaching the performance of digital processing. The multi-frequency approach can adapt to the frequency-selectivity of a simulated sub-6GHz scenario, attaining gains over frequency-flat digital processing. This constitutes the first demonstration that realistic MiLAC implementations can be configured to perform independent computations at different frequencies.

\end{abstract}
\begin{IEEEkeywords}
Microwave Linear Analog Computer (MiLAC), multiple-input single-output (MISO), wideband, LC resonators, frequency-selective.
\end{IEEEkeywords}

\section{Introduction}\label{section:intro}
\IEEEPARstart{T}{he} success of massive \ac{MIMO} \cite{marzetta, rusek} as a key enabling technology for 5G communications is driving future generation wireless communication systems towards increasing the number of antennas at the \ac{BS} \cite{emil_next,ultra_mMIMO, husha_data, xl-mimo, gMIMO}. This increase in the number of antennas allows for a refined spatial resolution and a higher beamforming gain, facilitating improvements in spectral efficiency \cite{husha_data,xl-mimo} and opening the door for higher frequency bands with large bandwidths available \cite{ultra_mMIMO,gMIMO}. 

Scaling conventional fully digital architectures, where each antenna is connected to a dedicated \ac{RF} chain, to the large-antenna regime is impractical due to the prohibitive requirements in terms of energy, cost, computational complexity, and interconnection bandwidth \cite{tri_hybrid, muris_pw, wax_journal}. To address this issue, \ac{MiLAC} has been recently proposed as an alternative processing architecture that shifts part of the multi-antenna processing to the analog domain \cite{milac,milac_gMIMO,milac_fund_lims, milac_cap,milac_coupling,milac_red_complx}. A \ac{MiLAC} is a multiport microwave network consisting of reconfigurable impedance components that can process signals injected at its input ports as they propagate through the network. This allows reducing the number of \ac{RF} chains in large multi-antenna systems without sacrificing performance, since it can implement optimal dimensionality-reduction/expansion entirely in the analog domain to match the number of spatial streams with the number of transmit/receive antennas \cite{milac,milac_gMIMO, milac_fund_lims, milac_cap,milac_coupling,milac_red_complx, unitary_wax}.

Initial works on \ac{MiLAC}-aided systems focused on narrowband scenarios where the impedance networks can adapt their response to the individual frequency considered \cite{milac,milac_gMIMO,milac_fund_lims, milac_cap,milac_coupling, milac_red_complx}. Under this narrowband assumption, \acp{MiLAC} are shown to be capacity-achieving for point-to-point \ac{MIMO} processing even under lossless-reciprocal constraints \cite{milac_cap}. These constraints are specially desirable since they ensure that no power is amplified or dissipated in the \ac{MiLAC}, reinforcing their energy-efficient operation. In \cite{milac_red_complx}, it is further shown that circuit complexity can be significantly reduced without affecting the capacity-achieving property. We will see in this work that this is only relevant for the narrowband case since the reduced \ac{DoFs} may otherwise be exploited towards facilitating wideband operation of \ac{MiLAC}. 

Despite the rapid progress in \ac{MiLAC}-aided processing, the main existing results are based on narrowband scenarios \cite{milac,milac_gMIMO,milac_fund_lims, milac_cap,milac_coupling, milac_red_complx}, while direct generalization to the wideband setting would require the over-idealistic assumption that impedance components can tune their response to individual narrowband frequency resources. Recent work in \cite{milac_ofdm} was the first to take into account the effect of frequency selectivity in a wideband OFDM system combining per-subcarrier digital processing with frequency-flat MiLAC processing. However, the assumption of having frequency-flat MiLAC processing via passive impedance networks is also an over-idealization, since impedance components as capacitors and inductors have an intrinsic frequency dependent behavior. Even assuming that perfect reconfiguration allows freely selecting inductance and/or capacitance of such components, integrating these components into a \ac{MiLAC} would still lead to non-trivial frequency-selective processing. This means that, even under frequency-flat channels, the capacity-achieving \ac{MiLAC} processing schemes derived in \cite{milac,milac_gMIMO,milac_cap,milac_red_complx,unitary_wax} may not be applicable to wideband systems due to this inherent frequency selectivity of the processing. Thus, the applicability of realistic \ac{MiLAC} processing to wideband scenarios remains widely unknown. 

Another open question in \ac{MiLAC} processing is how to provide implementations for such reconfigurable impedance networks that are both practical and flexible. In \cite{milac_phase_shift}, \ac{MiLAC} implementations are studied based on phase shifters and couplers. However, these implementations are shown to significantly constrain the processing capabilities of \ac{MiLAC}. Instead, reconfigurable inductors and capacitors \cite{var_ind1,var_ind3, varactor, var_ser_par} allow synthesizing any passive impedance network at given microwave frequencies \cite{pozar,circuits}, offering the full flexibility associated to narrowband \acp{MiLAC}. However, to the best of our knowledge, the available literature has not considered practical \ac{MiLAC} implementations based on variable inductors and/or capacitors.

This work addresses the previous gaps by studying a wideband \ac{MISO} scenario where the processing is performed through fully-connected \ac{MiLAC} implementations based on reconfigurable LC circuits. Specifically, we will consider implementations where each branch consists of a parallel or a series LC resonator, allowing for arbitrary impedances at given frequencies, but offering different frequency-dependent behavior.  Moreover, it is well known that reconfigurable capacitors, also termed varactors or varicaps, are easier to manufacture and integrate than variable inductors \cite{var_ind1,var_ind3, varactor},  so we will focus on \ac{MiLAC} implementations with fixed inductance and variable capacitance. Instead of minimizing the number of connections in the narrowband capacity-achieving \ac{MiLAC} as \cite{milac_red_complx}, we characterize the degrees of freedom available in the fully connected architecture after fixing the capacity-achieving constraint at a given reference frequency. We then explore how to exploit these degrees of freedom towards improving the wideband performance in both frequency-flat and frequency selective channels. The list of contributions is summarized as follows:
\begin{itemize}
\item We study the use of fully-connected \ac{MiLAC} implementations where each impedance interconnecting two ports (or one port to ground) corresponds to a parallel or series LC-resonator with reconfigurable capacitor. The map between the optimal MISO processing at a given frequency and the vector of capacitance values in these \ac{MiLAC} implementations is further characterized.
\item We prove that, after fixing optimal precoding at the reference frequency in a  \ac{MISO} system with $M$ transmit antennas, there are $\frac{M(M-1)}{2}+1$ \ac{DoFs} remaining to configure the \ac{MiLAC} processing. These \ac{DoFs}, initially derived in the space of \ac{MiLAC} scattering matrices at the reference frequency, are parameterized through an $(M-1)\times (M-1)$ unitary symmetric matrix and a unimodulus complex scalar. We further show that, when mapping back the \ac{DoFs} to the space of capacitor vectors, the full \ac{DoFs} survive as long as optimal processing is feasible within the constrained capacitance values set. This elucidates the potential of exploiting these \ac{DoFs} to attain wideband \ac{MiLAC} processing.
\item We propose a closed-form selection of the derived \ac{DoFs} to approach frequency-flat \ac{MiLAC} processing by avoiding the divergence of the susceptance values. This scheme is shown to be effective for \ac{MiLAC}-based processing under frequency-flat channels, further facilitating the feasibility of optimal \ac{MiLAC} processing at the reference frequency with parallel LC resonators having only capacitance reconfiguration. Note that parallel LC resonators with fixed inductance have a limited susceptance range, so that achieving optimal processing is not always possible, as will be elucidated throughout this work.
\item We show how to optimize the derived \ac{DoFs} to improve the \ac{MiLAC} processing performance at frequencies other that the reference one. This multi-frequency scheme is shown to be effective for \ac{MiLAC}-based processing under frequency-selective channels.
\end{itemize}
To the best of our knowledge, this work is the first to demonstrate that a realistic \ac{MiLAC} implementation can be configured to synthesize different processing filters at different frequencies without relying on digital processing. This strengthens the potential of this technology as an alternative to traditional digital processing.

The rest of the paper is organized as follows. In Section~\ref{sec:sys_mod}, the system model is presented, including the \ac{MISO} optimal processing under frequency-dependent behavior. Section~\ref{sec:milac_imp} presents the considered LC-resonator-based \ac{MiLAC} implementations, mapping the vector of reconfigurable capacitance with the resulting admittance matrix. In Section~\ref{sec:DoFs}, the \ac{DoFs} of a \ac{MiLAC} attaining optimal \ac{MISO} processing at a reference frequency are derived, and subsequently mapped to the \ac{DoFs} in the capacitance value selection for the considered \ac{MiLAC} implementations. Section~\ref{sec:freq_flat} provides a closed-form configuration of these \ac{DoFs} to reduce the frequency-sensitivity of the \ac{MiLAC} processing. Section~\ref{sec:multi_frec} shows how to instead use the \ac{DoFs} to improve the optimality of the \ac{MiLAC} processing at frequencies other than the reference frequency. In Section~\ref{sec:num_res}, numerical results are presented to confirm the validity of the theoretical derivations, as well as the performance of the proposed \ac{MiLAC} processing approaches in frequency-flat and frequency-selective \ac{MISO} scenarios. Section~\ref{sec:conc} concludes the paper with some final remarks and future work.

\subsection{Notation and Preliminaries}
In this paper, lowercase, bold lowercase and bold uppercase letters stand for scalars, column vectors and matrices, respectively. The operations $(\cdot)^{\text{T}}$, $(\cdot)^*$, $(\cdot)^{\text{H}}$, and $(\cdot)^{\dagger}$ denote transpose, conjugate, conjugate transpose, and pseudoinverse, respectively. The trace operator is written as $\mrm{tr}(\cdot)$, and the Euclidean norm as $\Vert\cdot\Vert$. The operation $\mathrm{diag}(\bs{a})$ gives a diagonal matrix with the input vector $\bs{a}$ as diagonal elements. $\bs{A} \otimes \bs{B}$ denotes the Kronecker product between matrices $\bs{A}$  and $\bs{B}$. $\mathbf{I}_i$ corresponds to the identity matrix of size $i$, $\boldsymbol{1}_{i\times j}$ denotes the $i \times j$ all-ones matrix,  and $\boldsymbol{0}_{i \times j}$ denotes the $i \times j$ all-zeros matrix (the subindices may be omitted if they can be derived from the context). $[\boldsymbol{A}]_{i,j}$ is the $(i,j)$th element of $\boldsymbol{A}$. Given an $N\times N$ matrix $\bs{X}$, its Cayley transform is defined as  \cite{cayley_prop}
\begin{equation}\label{eq:cayl_trans}
    \mrm{Cayley}(\bs{X})=(\mathbf{I}_N-\bs{X})(\mathbf{I}_N+\bs{X})^{-1}.
\end{equation} Given a scalar field $\mathcal{X}$, e.g., $\mathbb{R}$ or $\mathbb{C}$, $\mrm{Sym}_{\mathcal{X}}(N)\subset \mathcal{X}^{N\times N}$ denotes the respective set of symmetric matrices. $\mathcal{U}(N)$, $\mathcal{O}(N)$, and $\mathcal{U}_{\mrm{Sym}}(N)$ denote the space of $N\times N$ unitary, orthogonal, and unitary symmetric matrices, respectively. We write $\mathcal{A}\cong \mathcal{B}$ to denote that the sets $\mathcal{A}$ and $\mathcal{B}$ are diffeomorphic, i.e., there exists a continuously differentiable bijection between them whose inverse is also continuously differentiable.

\begin{defi}
    In the context of this work \textit{\ac{DoFs}} are defined as the number of independent real dimensions.
\end{defi}

From \cite{inverse_block} we have the following expression for the inverse of a $2\times 2$ block matrix
\begin{equation}\label{eq:inv_2x2}
    \begin{bmatrix}
        \bs{A} & \bs{B} \\ \bs{C} & \bs{D}
    \end{bmatrix}^{-1}= \begin{bmatrix}
        \bs{A}^{-1}+\bs{A}^{-1}\bs{B}\bs{E}^{-1}\bs{C} \bs{A}^{-1} & -\bs{A}^{-1}\bs{B}\bs{E}^{-1} \\ -\bs{E}^{-1}\bs{C}\bs{A}^{-1} & \bs{E}^{-1}
    \end{bmatrix},
\end{equation}
where the block $\bs{A}$ is invertible, and we have defined
\begin{equation}
    \bs{E} = \bs{D}-\bs{C}\bs{A}^{-1}\bs{B}.
\end{equation}

\section{System model}\label{sec:sys_mod}
We consider a \ac{MISO} scenario where an $M$-antenna device transmits a wideband signal to a single antenna receiver. The received signal may be expressed in frequency domain as
\begin{equation}
y(f)=\bs{h}^\mrm{T}(f)\bs{w}(f)s(f)+n(f),
\end{equation}
where $f$ denotes frequency, $\bs{h}(f)\in \mathbb{C}^{M\times 1}$ is the frequency dependent channel vector, $\bs{w}(f)\in \mathbb{C}^{M\times 1}$ is the frequency dependent precoding vector, with normalized frequency-flat power $\Vert\bs{w}(f)\Vert^2 =1$, $s(f)$ is the frequency spectrum of the transmitted signal, with power spectral density $\mathbb{E}\{\vert s(f)\vert^2\}=E_{s}(f)$, and $n(f)$ is the \ac{AWGN} term at frequency $f$, which is assumed to be \ac{IID} across different frequencies as $n(f)\sim \mathcal{CN}(0, N_0)$. The received \ac{SNR} at frequency $f$ is then given by
\begin{equation}\label{eq:snr}
    \mrm{SNR}(f) = \frac{E_{s}(f)\vert \bs{h}(f)^\mrm{T}\bs{w}(f)\vert^2}{N_0}.
\end{equation}
The optimal processing filter to maximize this \ac{SNR} expression at each frequency is given by the respective space-frequency \ac{MF}, associated to frequency-dependent \ac{MRT} \cite{mimo}
\begin{equation}
\bs{w}(f) = \mrm{e}^{\mrm{j}\varphi(f)}\frac{\bs{h}^{*}(f)}{\Vert\bs{h}(f)\Vert},    
\end{equation}
where the factor $\Vert\bs{h}(f)\Vert$ ensures that $\bs{w}(f)$ has normalized unit norm at each $f$, while $\varphi(f)$ corresponds to an arbitrary phase shift that has no impact on the resulting \ac{SNR} in \eqref{eq:snr}.\footnote{To allow for coherent decoding under arbitrary modulations, $\varphi(f)$ should be known at the receiver, but the transmitter and receiver may agree on an arbitrary preselection with minor overhead.}

Traditional \ac{OFDM}-based \ac{MIMO} systems with \ac{BB} digital processing employ sparse piloting schemes to estimate the channel only at certain subcarriers \cite{molisch}. The same processing would then be reused to process a range of subcarriers within the coherence bandwidth, $B_{\mrm{coh}}$. If we consider the channel estimate for a reference frequency $f_0$, the \ac{MRT} vector,
\begin{equation}\label{eq:mrc_f0}
\bs{w}(f_0) = \mrm{e}^{\mrm{j}\varphi(f_0)}\frac{\bs{h}^{*}(f_0)}{\Vert\bs{h}(f_0)\Vert},
\end{equation}
would thus be applied to all the subcarriers within the coherence band around $f_0$, $f\in [f_0-B_{\mrm{coh}},f_0+B_{\mrm{coh}}]$. In reality, there would still be some level of mismatch within this coherence band due to the smooth frequency variation of the channel, but this is often negligible.

In this work, we consider the use of a passive-lossless \ac{MiLAC}, connected to a single \ac{RF}-chain, to implement the precoding associated to $\bs{w}(f)$. In a narrowband \ac{MISO} scenario, achieving \eqref{eq:mrc_f0} is possible with such \ac{MiLAC}, since the passive-lossless restrictions still allow the multiplication of an arbitrary norm-1 vector at a given frequency \cite{milac_gMIMO,milac_fund_lims}. The problem arises when considering transmissions across multiple frequencies, where the restrictions of \ac{MiLAC} wideband processing remain widely unknown. Even if we assume that the channel is approximately constant within the coherence band, the implementation of a passive \ac{MiLAC} consists of impedance components whose response is variable over frequency. Hence, configuring the \ac{MiLAC} such that it performs \ac{MRT} processing \eqref{eq:mrc_f0} at $f_0$, does not give any performance guarantee for $f\in [f_0-B_{\mrm{coh}},f_0+B_{\mrm{coh}}]$ since a specific \ac{MiLAC} implementation may have an extremely narrowband response. 

In the following, we will study the frequency behavior of specific \ac{MiLAC} implementations based on LC resonators with reconfigurable capacitance, which may be realized using varactor diodes and fixed inductors. We will focus on the case where the \ac{MiLAC} is configured to attain \ac{MRT} at a reference frequency $f_0$, which allows connecting the current problem with the established literature on narrowband \ac{MiLAC}-based processing \cite{milac,milac_gMIMO,milac_fund_lims, milac_cap,milac_coupling,milac_red_complx}.

\section{MiLAC Resonator-based Implementations}\label{sec:milac_imp}
We consider a \ac{MiLAC} consisting of an $N$-port network, where we have $N=M+1$ total ports associated the 1 input and the $M$ outputs of the \ac{MISO} precoder. All ports are connected to each other, and to ground, via reconfigurable impedances, corresponding to the standard general fully-connected architecture considered in previous \ac{MiLAC} literature \cite{milac,milac_gMIMO,milac_fund_lims, milac_cap,milac_coupling}. We assume further a reciprocal lossless (reactive) network such that the admittance matrix is symmetric and purely imaginary. We can then express the $N\times N$ admittance matrix as
\begin{equation}\label{eq:sus}
    \bs{Y}= \mrm{j}\bs{B},
\end{equation}
where $\bs{B}\in \mrm{Sym}_{\mathbb{R}}(N)$ corresponds to the real symmetric susceptance matrix. For the considered fully-connected architecture, $\bs{B}$ can be expressed in terms of the branch susceptances $B_{\ell,n}$ as
\begin{equation}\label{eq:branch2mat}
    [\boldsymbol{B}]_{\ell,n}=\left\{ \begin{matrix}
        -B_{\ell,n}, &  \ell\neq n \\
        \sum_{k=1}^{N} B_{\ell,k}, & \ell=n 
    \end{matrix}\right.,
\end{equation}
where $B_{\ell,n}=B_{n, \ell}$ corresponds to the reconfigurable susceptance connecting ports $\ell$ and $n$ for $\ell\neq n$, and $B_{\ell, \ell}$ corresponds to the reconfigurable susceptance connecting port $\ell$ to ground. We can then extract the individual susceptance elements by considering the upper (or lower) triangular part of the resulting branch susceptance matrix, leading to the following branch susceptance vector
\begin{equation}\label{eq:sus_vec}
    \bs{b} \triangleq [B_{\ell,n}]_{1\leq\ell\leq n \leq N} \in \mathbb{R}^{\frac{N(N+1)}{2}\times 1}
\end{equation}
The following lemma elucidates the linear relation between the vectorized susceptance matrix and the branch susceptance vector.

\begin{lem}\label{lem:lin_admitt} The relation between the susceptance matrix \eqref{eq:branch2mat} and the branch susceptance vector \eqref{eq:sus_vec} corresponds to a linear map that may be expressed as
    \begin{equation}\label{eq:branch_sus}
        \vec(\bs{B}) = \bs{M} \bs{b},
    \end{equation}
where $\bs{M}$ is an $N^2 \times N(N+1)/2$ matrix given by
    \begin{equation}
        \bs{M} = (\bs{A}_{\mrm{in}}\otimes \bs{A}_{\mrm{in}}) \bs{Z}_{\mrm{D}},
    \end{equation}
with $\bs{Z}_{\mrm{D}}$ corresponding to a fixed $\big(\frac{N(N+1)}{2}\big)^2 \times \frac{N(N+1)}{2}$ matrix of 1s and 0s fulfilling $\vec^{-1}(\bs{Z}_{\mrm{D}}\bs{b})=\diag(\bs{b})$, and with $\bs{A}_{\mrm{in}}$ corresponding to the $N\times \frac{N(N+1)}{2}$ incidence matrix of the underlying directed graph associated to \eqref{eq:branch2mat} (removing the row associated to the ground node). In the considered architecture, the underlying graph corresponds to a fully-connected graph with $N+1$ nodes (including ground) where the direction is determined by the current direction convention.
\begin{proof}
The branch admittance matrix contains in its diagonal the individual admittance values at each branch of a multi-port circuit, while its off-diagonal elements are associated to potential mutual coupling admittances \cite[Chapter~10]{laplacian_admittance}. Its connection with the (node) admittance matrix via the incidence matrix, $\bs{A}_{\mrm{in}} $, is well understood (see e.g., \cite[Chapter~10]{laplacian_admittance}). Given \eqref{eq:sus} the branch susceptance matrix and (node) susceptance matrix are simply the imaginary part of the admittance counterparts. We can thus write 
\begin{equation}
\bs{B}=\bs{A}_{\mrm{in}} \bs{\bs{B}_\mrm{br}}\bs{A}^\mrm{T}_{\mrm{in}},
\end{equation}
where the branch susceptance matrix, $\bs{\bs{B}_\mrm{br}}$, is diagonal assuming negligible coupling between branches. Considering a suitable indexing of the branches associated to the columns of $\bs{A}_{\mrm{in}}$, we can write  $\bs{\bs{B}_\mrm{br}}=\diag(\bs{b})$. Using the property of the vectorization operator $\vec(\bs{A}\bs{X}\bs{C})=(\bs{C}^\mrm{T}\otimes\bs{A})\vec(\bs{X})$ leads to \eqref{eq:branch_sus}.
\end{proof}
\end{lem}

To be able to characterize the frequency behavior of \acp{MiLAC}, we need to consider specific implementations for which the frequency-dependency can be accurately modeled. Given the reciprocal-lossless network assumption, potential implementations based on lumped elements may only use inductors and capacitors. Specifically, we will consider implementations based on LC resonators, given that (lossless) microwave circuits can be equivalently represented through LC resonators \cite{pozar,circuits}. In this sense, we have two options for implementing each admittance component, using a series LC resonator or using a parallel LC resonator. Series resonators have the benefit that they can cover a wider range of admittance values at the cost of higher sensitivity to component tolerances, while parallel resonators allow for a finer resolution within a reduced admittance range. We will consider designs based on both types of resonators and compare their characteristics.

The reconfigurability of a \ac{MiLAC} may be achieved by considering voltage-controlled capacitors, e.g., varactor diodes \cite{varactor}. Note that reconfigurable inductors are also available, but their use is much less extended due to their additional challenges in terms of implementation and integration \cite{var_ind1,var_ind3,varactor}. We will focus on minimal designs where the \ac{DoFs} in terms of reconfigurability coincide with the associated \ac{DoFs} of the resulting admittance matrix. This means that we consider a single reconfigurable element per admittance branch, such that the number of controllable elements $\frac{N(N+1)}{2}$ coincides with the dimension of the respective (purely imaginary) impedance matrix. Given the reconfigurable capacitor $C_{\ell,n}$ associated to the branch connecting nodes $\ell$ and $n$ (or $\ell$ and ground for $n=\ell$), we may define the reconfigurable capacitance vector
\begin{equation}\label{eq:cap_vec}
    \bs{c} \triangleq [C]_{1\leq \ell \leq n \leq N} \in [C_{\min},C_{\max}]^{\frac{N(N+1)}{2}\times 1} \subset\mathbb{R}^{\frac{N(N+1)}{2}\times 1},
\end{equation}
where $C_{\min}$ and $C_{\max}$ give the range of capacitance values achievable by the varactor.

Next, we focus on specific MiLAC implementations based on LC resonators, showing the connection between reconfigurable capacitance and resulting frequency-dependent branch susceptance.
\subsection{Series LC Branches}
\begin{figure}[h]
\vspace{-2em}
    \centering
    \includegraphics[width=1\linewidth]{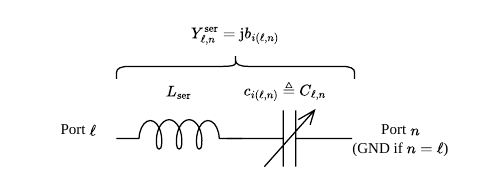}
    \caption{Schematic of series LC reconfigurable branch.}
    \label{fig:LC_ser}
\end{figure}

Let us consider a series LC resonator is used to connect nodes $\ell$ and $n$ (or $\ell$ to ground for $n=\ell$), corresponding to branch $i(\ell,n)$, as depicted in Fig.~\ref{fig:LC_ser}. Using standard circuit theory (see e.g., \cite{circuits}) we have that the $i$th branch susceptance is given by
\begin{equation}\label{eq:ser_B_res}
\begin{aligned}
    b_{i}(f,c_{i}) \triangleq g_{\mrm{ser},f}(c_i)= \frac{ 2\pi f c_{i}}{1-(2\pi f)^2 L_{\mrm{ser}}c_{i}}
\end{aligned}
\end{equation}
where, $L_{\mrm{ser}}$ is the fixed inductance and $c_i$ is the reconfigurable capacitance at branch $i\in\{1,\dots,\frac{N(N+1)}{2}\}$. Assuming a complete capacitance range,\footnote{Complete capacitance ranges may be practically approachable by considering switching networks made of parallel varactors, capacitors, and open/short circuits.} i.e., such that $C_{\min}=0$ and $C_{\max}=\infty$, the range of branch susceptance values is then restricted as
\begin{equation}\label{eq:ser_B_reg}
    b_i(f, c_i) \in \mathcal{B}_{\mrm{ser}} \triangleq \Big(-\infty,-\frac{1}{2\pi f L_{\mrm{ser}}}\Big)\cup [0, \infty ).
\end{equation}
Extending this range to be able to synthesize arbitrary susceptance values would then require selecting inductance values as large as possible, specifically $L_{\mrm{ser}}\gg\frac{1}{2\pi f}$.
\subsection{Parallel LC Branches}
\begin{figure}[h]
\vspace{-2em}
    \centering
    \includegraphics[width=1\linewidth]{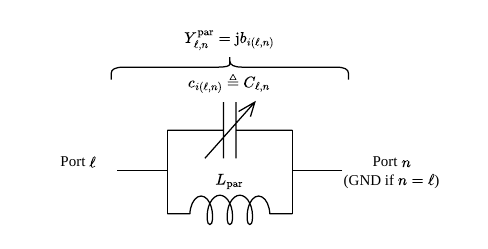}
    \caption{Schematic of parallel LC reconfigurable branch.}
    \label{fig:LC_par}
\end{figure}
Let us now consider a parallel LC resonator is used to connect nodes $\ell$ and $n$ (or $\ell$ to ground for $n=\ell$), corresponding to branch $i(\ell,n)$, as depicted in Fig.~\ref{fig:LC_par}. The resulting $i$th branch susceptance is then given by
\begin{equation}\label{eq:par_B_res}
\begin{aligned}
    b_{i}(f,c_i) \triangleq g_{\mrm{par},f}(c_i) =\frac{(2\pi f)^2 c_{i} L_{\mrm{par}}-1}{2\pi f L_{\mrm{par}}}
\end{aligned}
\end{equation}
where the variables have full correspondence with the series counterpart in \eqref{eq:ser_B_res}. In this case, considering a complete capacitance range we have the following restriction on the branch susceptance
\begin{equation}\label{eq:par_B_reg}
    b_i(f,c_i) \in\mathcal{B}_{\mrm{par}} \triangleq \Big[-\frac{1}{2\pi f L_{\mrm{par}}},\infty\Big).
\end{equation}
Extending this range to be able to synthesize arbitrary susceptance values would then require selecting inductance values as small as possible, specifically $L_{\mrm{par}}\ll\frac{1}{2\pi f}$.
However, for given $L_{\mrm{par}}>0$,  the absolute range of susceptance values is still significantly reduced with respect to \eqref{eq:ser_B_reg}. Instead, parallel branches have a monotonic susceptance relation in $c_i$, with a higher tolerance to small deviations in capacitance.

\subsection{Mixed LC Branches}
Alternative implementations may consider a combination of series and parallel LC branches. For example, we consider a fully-connected admittance network where the shunt branches correspond to parallel LC resonators and the series branches correspond to series LC resonators. This way, we may attain a high degree of reconfigurability within the signal path while maintaining a controlled shunt coupling to ground.

\section{Degrees of Freedom Characterization}\label{sec:DoFs}
In this section, we focus on deriving the \ac{DoFs} available in the \ac{MiLAC} after fixing optimal processing at $f_0$. We begin considering the scattering matrix domain, which corresponds to the domain where the processing \eqref{eq:mrc_f0} is attained. Subsequently, we will translate the \ac{DoFs} to the capacitance domain, associated to the physical reconfigurability of the \ac{MiLAC}.

\subsection{\ac{DoFs} in the Scattering Matrix Domain}
The frequency response of the MiLAC is fundamentally governed by the scattering matrix of the considered $N$-port network \cite{pozar,milac_fund_lims}. The connection between the scattering matrix and the admittance matrix, or susceptace matrix in lossless networks, is well understood from microwave network theory \cite[Chapter~4]{pozar}. Taking into account the symmetric constraint for reciprocal networks, this relation may be expressed through the Cayley transform as
\begin{equation}\label{eq:B2S}
    \bs{S}(f)= \mrm{Cayley}(\mrm{j}Z_0 \bs{B}(f)),
\end{equation}
where $Z_0$ is the characteristic impedance, typically set to $Z_0=50\Omega$. The lossless-reciprocal constraint from \eqref{eq:sus} translates under the Cayley transform to unitary symmetric constraints \cite{cayley_prop}, i.e., $\bs{S}(f)\in \mathcal{U}_\mrm{Sym}(N)$ where we have $\mathcal{U}_\mrm{Sym}(N)=\mathcal{U}(N)\cap \mrm{Sym}_\mathbb{C}(N)$. For the $N$-port network with $1$ input and $M$ outputs, we can write the scattering matrix in block form as
\begin{equation}\label{eq:block_S}
    \bs{S}(f)=\begin{bmatrix}s_{11}(f) & \bs{s}_{21}^\mrm{T}(f)\\\bs{s}_{21}(f) & \bs{S}_{22}(f) \end{bmatrix},
\end{equation}
where we have substituted $\bs{s}_{12}(f)=\bs{s}_{21}^\mrm{T}(f)$ considering the symmetric constraint. The input-output response of a \ac{MiLAC} is then given by $\bs{s}_{21}^\mrm{T}(f)$ \cite{milac_cap}, leading to
\begin{equation}\label{eq:w_to_s21}
    \bs{w}(f) = \bs{s}_{21}(f).
\end{equation}

The lossless-reciprocal constraint, $\bs{S}(f)\in \mathcal{U}_\mrm{Sym}(N)$, corresponds to simultaneously fulfilling $\bs{S}^\mrm{H}(f)\bs{S}(f)=\bs{S}(f)\bs{S}^\mrm{H}(f)=\mathbf{I}_{N}$ and $\bs{S}^\mrm{T}(f)=\bs{S}(f)$. Considering the block division in \eqref{eq:block_S}, we may rewrite it as
\begin{subequations}\label{eq:rest_S}
\begin{empheq}[left=\empheqlbrace]{align}
\bs{S}_{22}^\mrm{H}(f) \bs{S}_{22}(f) + \bs{s}_{21}^*(f)\bs{s}_{21}^\mrm{T}(f)&= \mathbf{I}_{M},\\
\bs{s}_{21}^\mrm{H}(f)\bs{S}_{22}(f)+s_{11}^*(f)\bs{s}_{21}^\mrm{T}(f)&=\bs{0}_{1\times M},\\
\label{eq:rest_s11_n}\Vert \bs{s}_{21}(f)\Vert^2+\vert s_{11}(f)\vert^2 &=1,\\
\bs{S}_{22}(f)&=\bs{S}_{22}(f)^\mrm{T},
\end{empheq}
\end{subequations}
Restricting the processing at $f_0$ to be optimum corresponds to imposing the extra restriction
\begin{equation}
    \bs{S}(f_0)=\bs{S}_{f_0},
\end{equation}
where $\bs{S}_{f_0}$ should be a matrix whose blocks, defined as in \eqref{eq:block_S}, fulfill the general restriction \eqref{eq:rest_S}, as well as
\begin{equation}\label{eq:s21_f0}
   \bs{s}_{21,f_0}(\varphi_0) =  \mrm{e}^{\mrm{j}\varphi_0}\frac{\bs{h}^{*}(f_0)}{\Vert\bs{h}(f_0)\Vert},
\end{equation}
associated to \eqref{eq:mrc_f0}. Note that $\varphi_0\in(-\pi,\pi]$ captures one of the available \ac{DoFs} in the selection of $\bs{S}_{f_0}$. The unit-norm constraint considered in \eqref{eq:s21_f0} requires to have $s_{11}=0$ in \eqref{eq:rest_s11_n}, which was further shown to be optimal in \cite{milac_cap}. Substituting in \eqref{eq:rest_S} and removing unnecessary terms leads to
\begin{subequations}\label{eq:S22_mrc_cons}
\begin{empheq}[left=\empheqlbrace]{align}
\label{eq:S22_comp_h}\bs{S}_{22,f_0}^\mrm{H} \bs{S}_{22,f_0} + \frac{\bs{h}(f_0)\bs{h}(f_0)^\mrm{H}}{\Vert \bs{h}(f_0) \Vert^2}&= \mathbf{I}_{M},\\
\label{eq:S11_orth_mrc}\bs{S}_{22,f_0}\bs{h}(f_0)&=\bs{0}_{M\times 1},\\
\label{eq:sym_S11}\bs{S}_{22,f_0}&=\bs{S}_{22,f_0}^\mrm{T},
\end{empheq}
\end{subequations}
We can note that the DoF associated to $\varphi_0$ in \eqref{eq:s21_f0} is cancelled in \eqref{eq:S22_mrc_cons}, and the remaining equations are thus associated to a constraint set in $\bs{S}_{22,f_0}$ capturing the remaining \ac{DoFs}. We next focus on characterizing these \ac{DoFs}.

Let us define the space of $\bs{S}_{22,f_0}$ matrices fulfilling \eqref{eq:S22_mrc_cons} as
\begin{equation}\label{eq:M_S22(f0)}
    \mathcal{S}_{22,f_0} = \{\bs{S}_{22,f_0}\in\mrm{Sym}_{\mathbb{C}}(M)\vert \bs{S}_{22,f_0}\text{ satisfies \eqref{eq:S22_mrc_cons}} \}
\end{equation}
The following proposition states the \ac{DoFs} available for selecting an arbitrary $\bs{S}_{22,f_0}$ within $\mathcal{S}_{22,f_0}$.

\begin{prop}\label{prop:DoFs_S22}
The number of \ac{DoFs} available for selecting $\bs{S}_{22,f_0}\in\mathcal{S}_{22,f_0}$, i.e., such that optimal processing can be achieved at $f_0$, is given by
\begin{equation}\label{eq:DoFs_S22}
    \mrm{DoF}_{\mathcal{S}_{22},f_0} \triangleq \dim(\mathcal{S}_{22,f_0}) = \frac{M(M-1)}{2}.
\end{equation}
Specifically, we have the following diffeomorphic relation
\begin{equation}\label{eq:S22_diff_Qs}
    \mathcal{S}_{22,f_0} \cong \mathcal{U}_{\mrm{Sym}}(M-1).
\end{equation}
The space $\mathcal{S}_{22,f_0}$ can be then parameterized by a unitary symmetric matrix $\bs{Q}_{\mrm{S}}\in\mathcal{U}_{\mrm{Sym}}(M-1)$ as
\begin{equation}\label{eq:S22_param}
    \bs{S}_{22,f_0}(\bs{Q}_{\mrm{S}}) = \widetilde{\bs{U}}^*\bs{Q}_\mrm{S}\widetilde{\bs{U}}^\mrm{H},
\end{equation}
where $\widetilde{\bs{U}}$ is a fixed semi-unitary matrix whose columns form an orthonormal basis for the column null-space of $\bs{h}(f_0)$.
\begin{proof}
See Appendix.
\end{proof}
\end{prop}

The previous proposition can be combined with the DoF associated to \eqref{eq:s21_f0} to derive the complete \ac{DoFs} available after fixing the optimal processing at $f_0$. To this end, let $\mathcal{S}_{f_0}$ denote the space of scattering matrices $\bs{S}_{f_0}$ attaining optimal processing at $f=f_0$, which is defined from the constraints \eqref{eq:s21_f0} and \eqref{eq:S22_mrc_cons}, with the block notation given in \eqref{eq:block_S}. Formally, we can write
\begin{equation}\label{eq:M_S(f0)}
    \mathcal{S}_{f_0} = \{\bs{S}_{f_0}\in\mathcal{U}_{\mrm{Sym}}(N)\vert \bs{S}_{f_0}\text{ satisfies \eqref{eq:s21_f0} and \eqref{eq:S22_mrc_cons}} \}
\end{equation}
We then have the following corollary derived from Proposition~\ref{prop:DoFs_S22}.
\begin{cor}\label{cor:DoFs_S}
The number of \ac{DoFs} available for selecting $\bs{S}_{f_0}\in\mathcal{S}_{f_0}$, i.e., such that optimal processing can be achieved at $f_0$, is given by
\begin{equation}\label{eq:DoFs_S}
    \mrm{DoF}_{\mathcal{S}_{f_0}}\triangleq \dim(\mathcal{S}_{f_0})=\frac{M(M-1)}{2}+1.
\end{equation}
Specifically, we have the following diffeomorphic relation
\begin{equation}\label{eq:S_space_diff}
    \mathcal{S}_{f_0}\cong \mathcal{U}(1)\times\mathcal{U}_\mrm{Sym}(M-1).
\end{equation} 
The \ac{DoFs} may be used to parameterize the scattering matrix  as $\bs{S}_{f_0}(\bs{Q}_{\mrm{S}},\varphi_0)$ by $\bs{Q}_{\mrm{S}}\in \mathcal{U}_{\mrm{Sym}}(M-1)$ and $\varphi_0\in(-\pi,\pi]$ according to \eqref{eq:S22_param} and \eqref{eq:s21_f0}, respectively.
\begin{proof}
    This result follows directly from the combination of the DoF associated to $\varphi_0$ in \eqref{eq:s21_f0} with those derived in Proposition~\ref{prop:DoFs_S22}, which were shown to be independent since they are associated to constraint sets on different blocks of $\bs{S}_{f_0}$ with no shared variables, as seen from \eqref{eq:s21_f0} and \eqref{eq:S22_mrc_cons}. The alternative parameterization of the DoF associated to $\varphi_0$ comes from the fact that this number shows up in \eqref{eq:s21_f0} within a uni-modulus complex exponential expression.
\end{proof}
\end{cor}

\begin{rmk}
An interesting observation is that, given the diffeomorphic relation from \eqref{eq:S_space_diff}, since $\mathcal{U}(1)$ and $\mathcal{U}_\mrm{Sym}(M-1)\cong\mathcal{U}(M-1)/\mathcal{O}(M-1)$ correspond to Riemannian manifolds when equipped with their Riemannian metric, we can also identify $\mathcal{S}_{f_0}$ with a Riemannian manifold by equipping it with the respective product metric. This may be useful in future work towards defining advanced optimization algorithms within this space.
\end{rmk}

\subsection{\Ac{DoFs} in the Capacitance Domain}
The \ac{DoFs} derived in Corollary~\ref{cor:DoFs_S} capture the freedom that we have in selecting a scattering matrix that attains optimal processing at a given frequency $f_0$. However, the reconfigurability of the considered MiLAC implementations is rather associated to the selection of the capacitance values of the respective resonator circuits. Hence, to be able to pratically expoit the \ac{DoFs} from Corollary~\ref{cor:DoFs_S} we need to map them back to the space of capacitor values parameterized by $\bs{c}$ in \eqref{eq:cap_vec}.

Given an arbitrary $\bs{S}_{f_0}\in\mathcal{S}_{f_0}$, we can obtain the associated susceptance matrix by inverting the converse transformation defined in \eqref{eq:B2S}. Exploiting again the symmetry of the input matrix, we get
\begin{equation}\label{eq:S_mat2Ad_mat}
\begin{aligned}
    \bs{B}_{f_0} = -\mrm{j} Y_0 \,\mrm{Cayley}(\bs{S}_{f_0}),
\end{aligned}
\end{equation}
where $Y_0=1/Z_0$ corresponds to the characteristic admittance. We may then obtain the respective branch susceptance vector by inverting the linear map defined in \eqref{eq:branch_sus}, giving
\begin{equation}
    \bs{b}_{f_0} = \bs{M}^{\dagger} \vec(\bs{B}_{f_0}),
\end{equation}
where $\bs{M}^{\dagger}$ corresponds to a left pseudoinverse of $\bs{M}$. Note that, since $\vec(\bs{B}_{f_0})$ must lie in the $\frac{N(N+1)}{2}$-dimensional subspace of $\mathbb{R}^{N^2}$ spanned by the linear map \eqref{eq:branch_sus}, associated to the symmetric constraints on $\bs{B}$, any left pseudoinverse definition will give the same branch susceptance vector $\bs{b}_{f_0}$.

Once the branch susceptances at $f_0$ have been characterized, the branch capacitances can be obtained by inverting at $f=f_0$ the function \eqref{eq:ser_B_res} or \eqref{eq:par_B_res} depending on if the corresponding branch is associated to a series or parallel branch, respectively. For branch $i\in\{1,\dots, \frac{N(N+1)}{2}\}$, the resulting capacitance value is related to the $i$th entry of $\bs{b}_{f_0}$, $b_{i,f_0}$, as
\begin{equation}
    c_{i,f_0}^{r} = g^{-1}_{r,f_0}(b_{i,f_0}),
\end{equation}
where $r\in \{\mrm{ser}, \mrm{par}\}$, and where the inverse functions are respectively given by
\begin{subequations}\label{eq:b2c}
    \begin{equation}\label{eq:ser_b2c}
        g^{-1}_{\mrm{ser},f_0}(b_{i,f_0}) = \frac{b_{i,f_0}}{2\pi f_{0}+(2\pi f_0)^2L_\mrm{ser} \,b_{i,f_0}},
    \end{equation}
    \begin{equation}\label{eq:par_b2c}
        g^{-1}_{\mrm{par},f_0}(b_{i,f_0}) = \frac{2 \pi f_0 L_{\mrm{par}}b_{i,f_0}+1}{(2\pi f_0)^2L_\mrm{par}}.
    \end{equation}
\end{subequations}
Altogether, we can write the branch capacitor values attaining optimal processing at $f_0$ as a function of the \ac{DoFs} derived in Corollary~\ref{cor:DoFs_S} as
\begin{equation}\label{eq:S-to-C}
c_{i,f_0}^{r}(\bs{Q}_{\mrm{S}},\varphi_0) = g^{-1}_{r,f_0}\!\!\left( \!\bs{M}^{\dagger} \vec\left(-\mrm{j}Y_0 \mrm{Cayley}\left(\bs{S}_{f_0}(\bs{Q}_{\mrm{S}},\varphi_0)\right)\right)\!\right),
\end{equation}
with $r\in\{\mrm{ser},\mrm{par}\}$ and $i\in \{1,\dots,\frac{N(N+1)}{2}\}$. Note that the result will generally lie in $\mathbb{R}^{\frac{N(N+1)}{2}\times 1}$, so we may still end up with non-valid solutions where $c_{i,f_0}^{r}(\bs{Q}_{\mrm{S}},\varphi_0)\notin [C_{min},C_{\max}]$ for some $i$. Even if we assume $C_{\min}=0$ and $C_{\max}\to \infty$, the required susceptance could result in a negative capacitance, which means that we would need inductive behavior to attain the desired response. These unfeasible results may be avoided by exploiting the \ac{DoFs} in selecting $\bs{Q}_{\mrm{S}}$ and $\varphi_0$, as will be explored in Section~\ref{sec:freq_flat}.

Let us use $\bs{c}_{f_0}(\cdot)$ to denote the general vector valued function mapping the scattering matrix domain $\mathcal{S}_{f_0}$ and the ambient capacitance domain, corresponding to the unconstrained higher dimensional space enclosing the capacitance values, in this case $\mathbb{R}^{\frac{N(N+1)}{2}\times 1}$. Specifically, each entry of $\bs{c}_{f_0}(\cdot)$ is selected from \eqref{eq:S-to-C}, which may be associated to either a series or parallel resonator. The following theorem formally states the \ac{DoFs} available in the capacitance domain after ensuring optimal processing at $f_0$.
\begin{thm}\label{thm:DoFs_C}
Assuming optimal \ac{MiLAC} processing \eqref{eq:mrc_f0} at $f_0$ is achievable at a strictly feasible regular point of $\bs{c}_{f_0}(\cdot)$, i.e., $\exists\bs{S}_{f_0}^{\star}\in\mathcal{S}_{f_0}$ s.t. $ \bs{c}_{f_0}(\bs{S}_{f_0}^{\star})\in (C_{\min},C_{\max})^{\frac{N(N+1)}{2}}$ with $c_{f_0,i}(\bs{S}_{f_0}^{\star})\neq \frac{1}{(2\pi f_0)^2 L_{\mrm{ser}}}$ for any series branch $i$, the number of \ac{DoFs} available for selecting reconfigurable capacitor values, $\bs{c}\in \bs{c}_{f_0}(\mathcal{S}_{f_0})\cap [C_{\min},C_{\max}]^{\frac{N(N+1)}{2}}$, is
\begin{equation}
    \mrm{DoF}_{\bs{c}} = \frac{M(M-1)}{2}+1.
\end{equation}
\begin{proof}
We begin by noticing that the space of capacitor values $\bs{c}\in [C_{\min},C_{\max}]^{\frac{N(N+1)}{2}\times 1}$ has equal real dimension, $\frac{N(N+1)}{2}$, as the space of scattering matrices, which under lossless reciprocal constraints corresponds to $\mathcal{U}_\mrm{Sym}(N)$. From \cite[Theorem~6]{cayley_prop}, the Cayley transform in \eqref{eq:S_mat2Ad_mat} is a diffeomorphism between the open dense subset of $\mathcal{U}_\mrm{Sym}(N)$ containing no eigenvalues at -1 and the space of purely imaginary symmetric matrices. From Lemma~\ref{lem:lin_admitt}, the map between node and branch admittances corresponds to a full-rank linear transformation, since it linearly maps a $\frac{N(N+1)}{2}$-dimensional subspace of $\mathbb{R}^{N^2}$, associated to the symmetric constraints, onto $\mathbb{R}^{\frac{N(N+1)}{2}}$. Hence, it trivially corresponds to a diffeomorphism. In the case of only parallel resonators, the map \eqref{eq:par_b2c} corresponds to a full-rank affine transformation, and thus a diffeomorphism. In the case of having some branches associated to series resonators, however, the map \eqref{eq:ser_b2c}, with inverse \eqref{eq:ser_B_res}, gives a diffeomorphism between the open dense subset of branch susceptance vectors where none of their entries associated to series resonators are poles of \eqref{eq:ser_b2c}, i.e., removing the points where $b_{i,f_0}=-\frac{1}{2\pi f_0 L_\mrm{ser}}$ for any series branch $i$, and the open dense subset of capacitance vectors where none of the entries associated to series resonators are poles of \eqref{eq:ser_B_res}, i.e., removing the points where $c_{i}=\frac{1}{(2\pi f_0)^2 L_\mrm{ser}}$ for any series branch $i$. Hence, the capacitance map $\bs{c}_{f_0}(\cdot)$ constitutes a diffeomorphism onto its image on its regular domain, defined by excluding the singularities of the Cayley transform, associated to infinite parallel capacitance and series capacitance $c_i=\frac{1}{(2\pi f_0)^2L_\mrm{ser}}$, and the singularities of \eqref{eq:ser_b2c} associated to infinite series capacitance. If there exists a regular point $\bs{S}_{f_0}^{\star}\in \mathcal{S}_{f_0}$ that further fulfills the feasibility restriction $\bs{c}_{f_0}(\bs{S}_{f_0}^{\star})\in (C_{\min},C_{\max})^{\frac{N(N+1)}{2}}$ (excluding the boundaries), continuity ensures that an open neighborhood of $\bs{S}_{f_0}^{\star}$ within $\mathcal{S}_{f_0}$ remains regular and feasible. Hence, the number of available \ac{DoFs} is preserved by the diffeomorphic capacitance map, which closes the proof.
\end{proof}
\end{thm}

Unlike the results from Corollary~\ref{cor:DoFs_S} within the scattering matrix domain, Theorem~\ref{thm:DoFs_C} does not necessarily imply that you can always find $\bs{c}$ attaining optimal processing at $f_0$ due to the global map between scattering matrices and capacitor values being only locally diffeomorphic. The formal requirement is then that the space of solutions achieving optimal processing at $f_0$, expressed in term of restrictions in the ambient capacitance domain, should intersect with the space of possible capacitance values from \eqref{eq:cap_vec} excluding singularities. The interpretation of Theorem~\ref{thm:DoFs_C} is then that, if such regular capacitance vector attaining optimal processing at $f_0$ exists, the resulting feasible set allows making small jumps in all directions within a space of dimension $\frac{M(M-1)}{2}+1$. The extent of these jumps such that the result is still contained within this feasible set is fundamentally limited by the intersection between the optimum processing set and the capacitance constraints set. As previously noted, assuming complete capacitance range, i.e., $C_{\min}=0$ and $C_{\max}\to\infty$, and inductance values satisfying $L_\mrm{ser}\gg \frac{1}{2\pi f}$ and/or $L_\mrm{par}\ll \frac{1}{2\pi f}$, the susceptance ranges in \eqref{eq:ser_B_reg} and/or \eqref{eq:par_B_reg} would approximately cover the full range of susceptances. Hence, the conditions for having the \ac{DoFs} in Theorem~\ref{thm:DoFs_C} would be satisfied by just avoiding the isolated points associated to singularities.

\section{Frequency-Flat MiLAC Processing}\label{sec:freq_flat}
In the previous section, we derived the scattering matrix restrictions attaining optimal processing at $f_0$, and mapped these to restrictions in the capacitance domain through the map defined through \eqref{eq:S-to-C}. In this section, we will explore how to exploit the \ac{DoFs} available to achieve frequency-flat \ac{MiLAC} processing.

\subsection{Connecting \ac{MiLAC} Frequency Selectivity with Susceptance}
The frequency selectivity of the \ac{MiLAC} processing can be identified with the derivative of the \ac{MiLAC} response with respect to frequency, measuring the variation to small changes in frequency. Since the \ac{MiLAC} processing corresponds to a block of the scattering matrix, let us consider the variability of the whole scattering matrix with respect to changes in frequency.  The frequency dependency in the considered \ac{MiLAC} implementations is connected to the branch susceptances through \eqref{eq:ser_B_res} and \eqref{eq:par_B_res}. Mapping this frequency dependency to the scattering domain is challenging, due to its level of entanglement across susceptance branches, as well as due to the involvement of matrix inverse operations in the Cayley transform. We will thus begin by connecting the frequency variability in the scattering domain with that in the susceptance domain. For presentation purposes, we consider derivatives with respect to angular frequency $\omega=2\pi f$. Moreover, we are interested on reducing the frequency selectivity around the reference frequency, associated with the evaluation of the derivatives at $\omega_0=2\pi f_0$, since this is the frequency at which the \ac{MiLAC} is configured for optimal processing.

Applying the chain rule to differentiate \eqref{eq:B2S} with respect to $\omega$, and substituting the derivative of the Cayley transform (see, e.g., \cite[Lemma~8.8]{cay_diff}), we get 
\begin{equation}\label{eq:diff_S}
    \frac{\mrm{d}\bs{S}(\omega)}{\mrm{d}\omega} \!\!=\!\! -\mrm{j} 2Z_0 (\bs{I}_{N}+\mrm{j}Z_0\bs{B}(\omega))^{-1}\frac{\mrm{d}\bs{B}(\omega)}{\mrm{d}\omega}(\bs{I}_{N}+\mrm{j}Z_0\bs{B}(\omega))^{-1}\!.
\end{equation}
It may be noted that $(\bs{I}_{N}+\mrm{j}Z_0\bs{B}(\omega))^{-1}$ is contained in a bounded subset of $\mathbb{C}^{N\times N}$, since its eigenvalues are given by $\frac{1}{1+\mrm{j}Z_0\lambda_{\bs{B}}(\omega)}$, where $\lambda_{\bs{B}}(\omega)$ are the real eigenvalues of the (real-symmetric) susceptance matrix, $\bs{B}(\omega)$, leading to $\frac{1}{\vert 1+\mrm{j}Z_0\lambda_{\bs{B}}(\omega)\vert}\leq 1$. The derivative of the susceptance matrix can be then connected to the branch susceptance derivatives through
\begin{equation}
    \frac{\mrm{d}\bs{B}(\omega)}{\mrm{d}\omega} = \vec^{-1}\Big(\bs{M}\frac{\mrm{d}\bs{b}(\omega)}{\mrm{d}\omega}\Big),
\end{equation}
where $\bs{M}$ is a fixed full-rank matrix with finite entries (as seen from Lemma~\ref{lem:lin_admitt}), and thus also bounded. We can then characterize the branch susceptance derivatives at $\omega_0$ for series and parallel branches by differentiating \eqref{eq:ser_B_res} and \eqref{eq:par_B_res}, respectively. For a series branch $i$ we then have
\begin{equation}\label{eq:diff_b_f_ser}
\begin{aligned}
    \left.\frac{\mrm{d}b_{i}(\omega)}{\mrm{d}\omega}\right\vert_{\omega_0} &= \frac{c_i(1-\omega_0^2L_\mrm{ser}c_i)+2\omega_0^2L_\mrm{ser}c_i^2}{(1-\omega_0^2L_\mrm{ser}c_i)^2}\\
    &\underset{\mrm{(a)}}{=}\frac{b_{i,f_0}}{\omega_0}+2L_\mrm{ser}b_{i,f_0}^2,
\end{aligned}
\end{equation}
where (a) comes from operating and substituting \eqref{eq:ser_B_res}, while for a parallel branch $i$ we have
\begin{equation}\label{eq:diff_b_f_par}
\begin{aligned}
    \left.\frac{\mrm{d}b_{i}(\omega)}{\mrm{d}\omega}\right\vert_{\omega_0} &= c_i+\frac{1}{\omega_0^2 L_\mrm{par}}\\
    &\underset{\mrm{(b)}}{=}\frac{b_{i,f_0}}{\omega_0}+\frac{2}{\omega_0^2 L_\mrm{par}}
\end{aligned}
\end{equation}
where (b) comes from substituting \eqref{eq:par_b2c}. If we want to minimize frequency selectivity at $\omega_0$, we could select the capacitance $c_i$ that minimizes \eqref{eq:diff_b_f_ser} or \eqref{eq:diff_b_f_par}, under series or parallel branch, respectively. However, this would lead to a specific selection of the capacitance vector, which is unlikely to attain optimal processing. Instead, we can note that the magnitude of both expressions depends on the magnitude of the susceptance, in the case of \eqref{eq:diff_b_f_ser} quadratically, and in the case of \eqref{eq:diff_b_f_par} linearly. Since the rest of the transformations leading to the derivative of the scattering matrix \eqref{eq:diff_S} are bounded, a suitable way to control frequency selectivity is to avoid large magnitudes in the susceptance values. The following lemma allows connecting the divergences in susceptance values with the singularities of the Cayley transform.
\begin{lem}\label{lem:inv_sing}
     Let $\bs{X}$ take values in a bounded subset $\mathcal{X}\subset \mathbb{C}^{N\times N}$. We then have
     \begin{equation}\label{eq:lem_eq}
        \max_{ 1\leq i,j\leq N} \big\vert [\bs{X}^{-1}]_{i,j}\big\vert\to \infty \iff \det(\bs{X})\to 0.
     \end{equation}
     \begin{proof}
Given the singular value decomposition $\bs{X}=\bs{U}\bs{S}\bs{V}^\mrm{H}$, we have that $\bs{X}^{-1}=\bs{V}\bs{S}^{-1} \bs{U}^\mrm{H}$, such that the singular values of $\bs{X}^{-1}$ are the inverse of the singular values of $\bs{X}$. For the forward implication "$\Rightarrow$" we assume that $\max_{ 1\leq i,j\leq N} \big\vert [\bs{X}^{-1}]_{i,j}\big\vert\to \infty$. We then have that $\sigma_{\max}(\bs{X}^{-1})\geq \max_{ 1\leq i,j\leq N} \big\vert [\bs{X}^{-1}]_{i,j}\big\vert \to\infty$, so that $\sigma_{\min}(\bs{X})=\frac{1}{\sigma_{\max}(\bs{X}^{-1})}\to 0$. Since the absolute value of the determinant is the product of all singular values, which are bounded in $\bs{X}$ from being in a bounded subset of $\mathbb{C}^{N\times N}$, we have that $\vert \det(\bs{X})\vert\to 0$. For the backward implication "$\Leftarrow$" we assume that $\det(\bs{X})\to 0$, so that $\sigma_{\min}(\bs{X})\to 0$. Since $\Vert \bs{X}^{-1}\Vert_2=\frac{1}{\sigma_{\min}(\bs{X})}$, we have that $\Vert \bs{X}^{-1}\Vert_2 \to \infty$. Hence, $\max_{i,j}\vert [\bs{X}^{-1}]_{i,j}\big\vert \geq \frac{1}{N}\Vert {\bs{X}^{-1}}\Vert_2\to \infty$, which concludes the proof.

     \end{proof}
\end{lem}

Lemma~\ref{lem:inv_sing} confirms that the points where one or more susceptance values diverge are associated with the singularities of the Cayley transform in \eqref{eq:S-to-C}. Note that the singularity set of the Cayley transform corresponds to the set where $(\mathbf{I}_{N}+\bs{S}_{f_0})$ in \eqref{eq:cayl_trans} is not invertible, i.e., has determinant 0. Moreover, the input to the Cayley transform in \eqref{eq:S_mat2Ad_mat} are scattering matrices, which are contained in a bounded subset of $\mathbb{C}^{N\times N}$ due to their unitary constraints. Thus,  $(\mathbf{I}_{N}+\bs{S}_{f_0})$ is also in a bounded subset, so that the inverse term in the Cayley transform would fulfill the left-hand side of \eqref{eq:lem_eq} whenever $b_{i,f_0}\to \pm \infty$ for some $i$. Since the other factor in the Cayley transform, $(\mathbf{I}_{N}-\bs{S}_{f_0})$, is also bounded, it cannot be the source of a diverging susceptance value, which would thus be associated with a divergence in $(\mathbf{I}_{N}+\bs{S}_{f_0})^{-1}$. Hence, reducing the frequency selectivity of the \ac{MiLAC} processing around $f_0$ can be associated with avoiding the singularities of the Cayley transform. This gives a tractable objective for exploiting the \ac{DoFs} available after fixing optimal processing at $f_0$, which admit a useful parametrization in the scattering matrix domain as seen in Theorem~1.

\begin{rmk}\label{rmk:frec_flat}
    A physical interpretation of the relation between frequency-selective behavior and diverging susceptance is that, when susceptance approaches infinity impedance approaches zero, due to their inverse relation. This can be associated with having some components at series resonance. The series resonance behavior is known to be associated with an extremely narrowband frequency transfer function (see, e.g., \cite[Chapter~6]{pozar}), specially assuming lossless circuits where the null-resistance would drive the Q-factor towards infinity. This reasoning suggest that, if we want to avoid a narrowband \ac{MiLAC} frequency response, we should avoid diverging susceptance values.
\end{rmk}

\begin{rmk}
    Another motivation to avoid divergent susceptance values is to improve the feasibility of optimal \ac{MiLAC} processing with parallel resonators, since the range of possible susceptance values (assuming only capacitance reconfiguration) is bounded from below, as seen in \eqref{eq:par_B_reg}. 
\end{rmk}

\subsection{DoFs Selection for Non-Singular Cayley Transform}
The divergent susceptance values, leading to high-frequency sensitivity of the \ac{MiLAC} processing, have been matched to the singularity set of the Cayley transform. We next provide a closed-form approach to select $\bs{Q}_\mrm{S}$ and $\varphi_0$ such that the input $\bs{S}_{f_0}\in \mathcal{S}_{f_0}$ to the Cayley transform in \eqref{eq:S_mat2Ad_mat} is as far away as possible from this singularity set. As previously mentioned, the singularities of the Cayley transform are associated to the term $(\mathbf{I}_{N}+\bs{S}_{f_0})^{-1}$. Applying the $2\times 2$ block inverse \eqref{eq:inv_2x2} with the blocks defined as in \eqref{eq:block_S}, and taking into account the identity
\begin{equation}\begin{aligned}
    \bs{S}_{22,f_0}+& \mathbf{I}_{M}-\bs{s}_{21,f_0}\bs{s}_{21,f_0}^\mrm{T}\\
    &=\bs{U}^{*}\left(\begin{bmatrix}
        \bs{Q}_\mrm{S} & \bs{0}_{(M-1)\times 1} \\ \bs{0}_{1\times (M-1)} &-e^{\mrm{j} 2 \varphi_0}
    \end{bmatrix} +\bs{U}^\mrm{T}\bs{U}\right)\bs{U}^\mrm{H},
\end{aligned}
\end{equation}
which can be derived from \eqref{eq:S22_s21_unit}, where $\bs{U}=[\widetilde{\bs{U}}, \bs{u}_M]$ with $\bs{u}_M=\frac{\bs{h}(f_0)}{\Vert\bs{h}(f_0)\Vert}$ and $\widetilde{\bs{U}}$ defined in Proposition~\ref{prop:DoFs_S22}, leads to
\begin{equation}
    \big(\mathbf{I}_{N}+\bs{S}_{f_0}\big)^{-1} = \begin{bmatrix}
        1+e^{\mrm{j}2\varphi_0}[\bs{\Theta}]_{M,M} & -e^{\mrm{j}\varphi_0}[\bs{\Theta}]_{M,:}\bs{U}^\mrm{T}\\
        -e^{\mrm{j}\varphi_0}\bs{U}[\bs{\Theta}]_{:,M} & \bs{U}\bs{\Theta}\bs{U}^\mrm{T}
    \end{bmatrix}.
\end{equation}
The singularity is now associated with the term $\bs{\Theta}$ given by
\begin{equation}
\begin{aligned}
    \bs{\Theta}(\bs{Q}_\mrm{S}, \varphi_0) &= \left(\begin{bmatrix}
        \bs{Q}_\mrm{S} & \bs{0}_{1\times (M-1)}\\ \bs{0}_{(M-1)\times 1} & -e^{\mrm{j}2\varphi_0}
    \end{bmatrix}+\bs{U}^\mrm{T}\bs{U}\right)^{-1} \\
    &= \left(\begin{bmatrix}
        \bs{Q}_\mrm{S}+\widetilde{\bs{U}}^\mrm{T}\widetilde{\bs{U}} & \widetilde{\bs{U}}^\mrm{T} \bs{u}_{M}\\ \bs{u}_{M}^\mrm{T}\widetilde{\bs{U}} & \bs{u}_{M}^\mrm{T}\bs{u}_{M}-e^{\mrm{j}2\varphi_0}
    \end{bmatrix}\right)^{-1}\\
    &=\begin{bmatrix}
        \bs{\Theta}_{11} & \bs{\theta}_{12}\\ \bs{\theta}_{12}^\mrm{T} & \theta_{22}
    \end{bmatrix}.
\end{aligned}
\end{equation}
A subsequent application of \eqref{eq:inv_2x2}, considering now its permuted version such that $\bs{A}=\bs{u}_{M}^\mrm{T}\bs{u}_{M}-e^{\mrm{j}2\varphi_0}$, $\bs{D}=\bs{Q}_\mrm{S}+\widetilde{\bs{U}}^\mrm{T}\widetilde{\bs{U}}$, and $\bs{C}=\bs{B}^\mrm{T}=\widetilde{\bs{U}}^\mrm{T} \bs{u}_{M}$, gives
\begin{subequations}\label{eq:Theta_sing}
\begin{equation}
\bs{\Theta}_{11}(\bs{Q}_\mrm{S},\bs{\varphi}_0) = \left(\bs{Q}_\mrm{S}+\widetilde{\bs{U}}^\mrm{T} \Big(\mathbf{I}_{M}-\frac{\bs{u}_{M}\bs{u}_M^\mrm{T}}{\bs{u}_M^\mrm{T}\bs{u}_M-e^{\mrm{j}2\varphi_0}}\Big)\widetilde{\bs{U}} \right)^{-1},
\end{equation}
\begin{equation}
    \bs{\theta}_{12}(\bs{Q}_\mrm{S},\bs{\varphi}_0) = -\frac{\bs{\Theta}_{11}(\bs{Q}_\mrm{S},\bs{\varphi}_0)\widetilde{\bs{U}}^\mrm{T}\bs{u}_{M}}{\bs{u}_M^\mrm{T}\bs{u}_M-e^{\mrm{j}2\varphi_0}} 
\end{equation}
\begin{equation}
    \theta_{22} = \frac{1}{\bs{u}_M^\mrm{T}\bs{u}_M-e^{\mrm{j}2\varphi_0}}+\frac{\bs{u}_M^\mrm{T}\widetilde{\bs{U}}\bs{\Theta}_{11}(\bs{Q}_\mrm{S},\bs{\varphi}_0)\widetilde{\bs{U}}^\mrm{T}\bs{u}_M}{(\bs{u}_M^\mrm{T}\bs{u}_M-e^{\mrm{j}2\varphi_0})^2}
\end{equation}
\end{subequations}
From \eqref{eq:Theta_sing}, it can be noted that the singularity set is associated to the restrictions
\begin{equation}\label{eq:sing_varphi}
    \bs{u}_M^\mrm{T}\bs{u}_M-e^{\mrm{j}2\varphi_0}=0,
\end{equation}
and
\begin{equation}\label{eq:sing_Qs}
\det\left(\bs{Q}_\mrm{S}+\widetilde{\bs{U}}^\mrm{T} \Big(\mathbf{I}_{M}-\frac{\bs{u}_{M}\bs{u}_M^\mrm{T}}{\bs{u}_M^\mrm{T}\bs{u}_M-e^{\mrm{j}2\varphi_0}}\Big)\widetilde{\bs{U}} \right)=0.
\end{equation}
It can be then noted that the left hand side of \eqref{eq:sing_varphi} corresponds to a sum of two complex numbers, where one is solely dependent on $\varphi_0$. Hence, a straightforward way to avoid the respective singularity is to sum both complex numbers in phase, such that the result is furthest away from $0$. This leads to the following selection of $\varphi_0$:
\begin{equation}\label{eq:varphi_fflat}
    \varphi_0^{\mrm{NS}} = \frac{\angle(-\bs{u}_M^\mrm{T}\bs{u}_M)}{2}.
\end{equation}
Note that this selection will always avoid the singularity associated to \eqref{eq:sing_varphi} since $\vert \bs{u}_M^\mrm{T}\bs{u}_M-e^{\mrm{j}2\varphi^{\mrm{NS}}_0}\vert\geq 1$. We can now substitute \eqref{eq:varphi_fflat} in \eqref{eq:sing_Qs}, and consider the Takagi factorization
\begin{equation}
\bs{U}_{\mrm{Tak}}\bs{\Sigma}_{\mrm{Tak}}\bs{U}_{\mrm{Tak}}^\mrm{T}= \widetilde{\bs{U}}^\mrm{T} \Big(\mathbf{I}_{M}-\frac{\bs{u}_{M}\bs{u}_M^\mrm{T}}{\bs{u}_M^\mrm{T}\bs{u}_M-e^{\mrm{j}2\varphi^{\mrm{NS}}_0}}\Big)\widetilde{\bs{U}}.
\end{equation}
Note that The remaining singularity is then associated to
\begin{equation}\label{eq:sing_QS_tak}
\det(\bs{Q}_\mrm{S}+\bs{U}_{\mrm{Tak}}\bs{\Sigma}_{\mrm{Tak}}\bs{U}_{\mrm{Tak}}^\mrm{T})=0.
\end{equation}
Due to the symmetric restriction of the summands within the determinant in \eqref{eq:sing_QS_tak}, the expression will only evaluate to 0 whenever the minimum singular value of the input evaluates to 0, i.e., $\sigma_{\min}(\bs{Q}_\mrm{S}+\bs{U}_{\mrm{Tak}}\bs{\Sigma}_{\mrm{Tak}}\bs{U}_{\mrm{Tak}}^\mrm{T})=0$. In order to avoid that, we can maximize the resulting singular values so that they are pushed away from the singularity set. Since the singular values of $\bs{Q}_\mrm{S}$ are restricted to $1$, considering Weyl's inequality for the singular value setting \cite[Corollary~7.3.5, (7.3.13)]{mat_anal} yields
\begin{equation}
    \sigma_{i}(\bs{Q}_\mrm{S}+\bs{U}_{\mrm{Tak}}\bs{\Sigma}_{\mrm{Tak}}\bs{U}_{\mrm{Tak}}^\mrm{T})\leq 1+[\bs{\Sigma}_{\mrm{Tak}}]_{i,i},
\end{equation}
where equality is achieved by selecting $\bs{Q}_\mrm{S}$ as
\begin{equation}\label{eq:QS_fflat}
    \bs{Q}_\mrm{S}^{\mrm{NS}} = \bs{U}_{\mrm{Tak}}\bs{U}_{\mrm{Tak}}^\mrm{T}.
\end{equation}
Note that this selection will always ensure that the singularity associated to \eqref{eq:sing_Qs} is avoided since the resulting minimum singular value will fulfill $\sigma_{\min}(\bs{Q}_\mrm{S}+\bs{U}_{\mrm{Tak}}\bs{\Sigma}_{\mrm{Tak}}\bs{U}_{\mrm{Tak}}^\mrm{T})\geq 1$.

\section{Multi-Frequency MiLAC Processing}\label{sec:multi_frec}
In the previous section, we showed how to exploit the \ac{DoFs} from the space of scattering matrices achieving optimal processing at $f_0$, namely $\mathcal{S}_{f_0}$, to attain quasi-frequency-flat \ac{MiLAC} processing. However, in some cases, the processing that should be applied across different frequency points may change rapidly, e.g., in rich multipath environments. In digital processing architectures, such cases may be handled by considering a finer frequency grid for applying processing across the spectrum. In the case of analog processing architectures, as \ac{MiLAC}, it is not clear whether the processing can be suitably adapted to a finer frequency grid. In this section, we will explore how to solve this problem by making use of the \ac{DoFs} derived in Section~\ref{sec:DoFs}.

Assume that we want to attain optimal processing within a set of frequencies $\{f_0, f_1, \dots, f_K \}$. Each frequency is associated to a constraint set given by \eqref{eq:s21_f0} and \eqref{eq:S22_mrc_cons} for the evaluation of the scattering matrix at $f_k$. Reusing the notation from \eqref{eq:M_S(f0)}, we denote the space of scattering matrices attaining optimal processing at $f_k$ as $\mathcal{S}_{f_k}$. From Corollary~\ref{cor:DoFs_S}, we know that each $\mathcal{S}_{f_k}$ corresponds to a submanifold of $\mathcal{U}_{\mrm{Sym}}(N)$ of dimension $\frac{M(M-1)}{2}+1$. Notably, if $\bs{h}(f_{i})=\bs{h}(f_{j})$ for some $(f_{i},f_{j})$, their respective optimal processing submanifolds would coincide in the scattering matrix domain, i.e., $\mathcal{S}_{f_i}=\mathcal{S}_{f_j}$.  However, each of these submanifolds may map to distinct manifolds in the ambient capacitance domain due to the frequency dependency of the inverse maps $g_{r,f}^{-1}(\cdot)$ in \eqref{eq:b2c}. We thus need to connect our optimal processing submanifolds within the ambient capacitance domain, where the magnitudes would indeed correspond to the same physical entity independent of the frequency. Achieving optimal processing for all the frequencies in the set $\{f_0, f_1, \dots, f_K \}$ would then correspond to finding a capacitance vector $\bs{c}$ within the set
\begin{equation}\label{eq:mult_freq_set}
    \mathcal{C}^{\mrm{Opt}}\triangleq \bigcap_{k=0}^{K}\bs{c}(\mathcal{S}_{f_k}) \cap [C_{\min},C_{\max}]^{\frac{N(N+1)}{2}}.
\end{equation}
A complete characterization of the set $\mathcal{C}^{\mrm{Opt}}$, including the conditions under which this set is non-empty, is beyond the scope of this work, and may be considered in future work. We will thus focus on practical approaches that still allow to demonstrate the potential of multi-frequency analog processing using the theory previously introduced.

Let us consider again that the \ac{MiLAC} processing has been selected to attain optimal at the reference frequency $f_0$. We can still use the \ac{DoFs} available in $\mathcal{S}_{f_0}$ to optimize the processing applied at the additional set frequencies $\{f_1,\dots, f_K\}$. In the context of \ac{MISO} processing, this leads to an optimization problem of the form\footnote{Alternative objective functions, such as $\max \min$ \ac{SNR} across the frequencies, or sum-\ac{SNR} in $\log$ scale, may also be of interest depending on the application. However, many of these will have the same solution assuming $\mathcal{C}^{\mrm{Opt}}$ is non-empty in \eqref{eq:mult_freq_set}.
}
\begin{equation}\label{eq:opt_prob}
\begin{aligned}
    \max_{\bs{Q}_\mrm{S}, \varphi_0} \sum_{k=0}^{K}& \;\mrm{SNR}(f_k,\bs{c}(\bs{Q}_{\mrm{S}},\varphi_0))\\
    \mrm{s.t. }& \bs{Q}_{\mrm{S}}\in \mathcal{U}_{\mrm{Sym}}(M-1),  
\end{aligned}
\end{equation}
where $\mrm{SNR}(f_k,\bs{c})$ corresponds to the \ac{SNR} at frequency $f_k$ when configuring the \ac{MiLAC} with a capacitance vector $\bs{c}$.\footnote{$\mrm{SNR}(f_k,\bs{c})$ can be obtained using \eqref{eq:snr}, where $\bs{w}(f)$, related to the scattering matrix through \eqref{eq:w_to_s21}, can be obtained through the composition of functions \eqref{eq:B2S}, \eqref{eq:sus_vec}, and \eqref{eq:ser_B_res} or \eqref{eq:par_B_res} under series or parallel branches, respectively.} Note that, if $\mathcal{C}^{\mrm{Opt}}$ is not empty, the solution to \eqref{eq:opt_prob} would coincide with this set, since the performance at the considered frequencies is limited by that achieved through optimal processing. The need to include the frequency $f_0$ in the sum evaluation comes from the previously discussed concern that not all points in $\mathcal{S}_{f_0}$ (parameterized through $\bs{Q}_\mrm{S}$ and $\varphi_0$) may lead to valid capacitor values when considering the restrictions in \eqref{eq:cap_vec}.

The optimization problem in \eqref{eq:opt_prob} is non-convex due to the unitary symmetric restrictions on $\bs{Q}_{\mrm{S}}$, as well as the non-linear coupling between \ac{DoFs} and objective function, which may require an independent study. We will focus on numerical optimization techniques to attain approximate solutions with the aim at showcasing the potential of the theory previously defined. In this sense, the main challenge lies in the unitary symmetric constraints associated to $\bs{Q}_{\mrm{S}}$, which limit the ability to numerically explore the associated space. The following proposition provides a convenient parametrization for $\mathcal{U}_{\mrm{Sym}}(N)$.
\begin{prop}\label{prop:Unit_Sym_dep}
The space of unitary symmetric matrices can be parameterized through bounded real vectors of dimension $N(N+1)/2$. Specifically, $\forall \bs{Q}_\mrm{S}\in \mathcal{U}_\mrm{Sym}(N)$ there exists $\bs{v}\in\big[-\sqrt{N}\pi,\sqrt{N}\pi\big]^{\frac{N(N+1)}{2}\times 1}$ such that
    \begin{equation}\label{eq:dep_UnitSym}
        \bs{Q}_\mrm{S}(\bs{v}) = \exp\big(\vec^{-1}(\bs{U}_{\mrm{G-M}}^{\mrm{Sym}}\bs{v})\big),
    \end{equation}
    where $\exp(\cdot)$ denotes the matrix exponential map, and $\bs{U}_{\mrm{G-M}}^{\mrm{Sym}}$ corresponds to a unitary matrix whose columns correspond to the vectorized and normalized entries of the subset of generalized Gell-Mann matrices scaled by the imaginary unit $\mrm{j}$ fulfilling symmetric constraints \cite{gen_gell_mann}. Specifically, $\bs{U}_{\mrm{G-M}}^{\mrm{Sym}}$ can be constructed from the vectorization of the first $N(N+1)/2$ basis elements defined in \cite[Eq. (10)]{dim_eff_param}.
\begin{proof}
    This proposition is linked to the dimensionally-efficient parametrization studied in \cite{dim_eff_param} for the unitary group. As noted in \cite{dim_eff_param}, the generalized Gell-Mann matrices scaled by the imaginary unit $\mrm{j}$ form a basis for the space of skew-Hermitian matrices, which can be mapped to the space of unitary matrices through the matrix exponential. The subset of $N(N+1)/2$ symmetric matrices out of these scaled generalized Gell-Mann matrices, i.e., the first $N(N+1)/2$ elements in \cite[Eq. (10)]{dim_eff_param} with purely imaginary entries,  form a basis for the space of skew-Hermitian matrices under additional symmetry constraint. Note that the basis elements are orthogonal to each other and the number of basis elements matches the dimension of this skew-Hermitian symmetric space. From the power series expansion of the matrix exponential, it can be trivially noted that the symmetry restriction is maintained at the output, mapping the skew-Hermitian symmetric space to the unitary symmetric space. The boundedness of the entries of $\bs{v}$ then falls from \cite[Proposition~2]{dim_eff_param}, where we can see that the symmetry restriction corresponds to having some parametrization entries fixed to 0, i.e., those associated to non-symmetric basis elements in \cite[Eq. (10)]{dim_eff_param}. 
\end{proof}
\end{prop}

The parametrization defined in Proposition~\ref{prop:Unit_Sym_dep}, applied to $\mathcal{U}_\mrm{Sym}(M-1)$, is associated to a compact subset of $\mathbb{R}^{\frac{M(M-1)}{2}\times 1}$. This may be convenient when considering exhaustive search approaches since the limits of the search are bounded. The following corollary allows to relax the compactness of the parametrization such that unconstrained optimization approaches are also applicable.
\begin{cor}\label{cor:unc_param}
  Relaxing the parametrization domain in \eqref{eq:dep_UnitSym} to $\mathbb{R}^{\frac{N(N+1)}{2}\times 1}$ has no impact on the resulting image, i.e., $\forall \bs{v}\in \mathbb{R}^{\frac{N(N+1)}{2}\times 1}$ we have
  \begin{equation}
    \bs{Q}_\mrm{S}(\bs{v})\in \mathcal{U}_{\mrm{Sym}}(N).
  \end{equation}
  \begin{proof}
      The matrix exponential in \eqref{eq:dep_UnitSym} can be defined through the complex exponential of the eigenvalues of $\vec^{-1}(\bs{U}_{\mrm{G-M}}^{\mrm{Sym}}\bs{v})$, which are purely imaginary due to the skew-Hermitian restriction, as noted in \cite{dim_eff_param}. The periodicity of the complex exponential thus constrains the output eigenvalues to uni-modulus, keeping the resulting matrices within the unitary group. The symmetric structure is still enforced by the respective subset of basis elements, as seen from the proof of Proposition~\ref{prop:Unit_Sym_dep}.
  \end{proof}   
\end{cor}\label{rmk:rel_R}

Altogether, by considering the parametrization defined in Proposition~\ref{prop:Unit_Sym_dep}, with the considerations from Corollary~\ref{cor:unc_param}, we may rewrite the optimization from \eqref{eq:opt_prob} as an unconstrained optimization problem of the form 
\begin{equation}\label{eq:opt_prob_unc}
    \max_{\bs{v}, \varphi_0} \sum_{k=0}^{K} \;\mrm{SNR}(f_k,\bs{c}(\bs{Q}_{\mrm{S}}(\bs{v}),\varphi_0)),  
\end{equation}
where $\bs{v}\in \mathbb{R}^{\frac{M(M-1)}{2}\times 1}$. In the next section, we will showcase the potential of exploiting the \ac{DoFs} derived in Section~\ref{sec:DoFs} to perform multi-frequency processing by solving \eqref{eq:opt_prob_unc} using numerical optimization tools.

\section{Numerical results}\label{sec:num_res}
Let us consider the standard Saleh-Valenzuela channel model \cite{saleh_val} applied to the \ac{MISO} setting with a half-wavelength spaced uniform linear array at the transmitter.\footnote{The \ac{MISO} channel considerations are motivated by the good compromise between practicality, generality, and interpretability offered by such model, but the main findings of this work are largely independent of the specific setting.} The corresponding \ac{MISO} channel is given by
\begin{equation}\label{eq:sal_val_ch}
    \bs{h}^\mrm{T}(f) = \sum_{i=1}^{N_\mrm{MP}}\beta_i e^{-\mrm{j}2\pi f\tau_i}\bs{a}^\mrm{T}(f,\phi_i),
\end{equation}
where $\beta_i$, $\tau_i$, and $\phi_i$ are the complex gain, delay, and angle of departure, respectively, for the $i$th path, $N_\mrm{MP}$ is the number of multi-path components, and $\bs{a}^\mrm{T}(f,\phi_i)$ is the far-field frequency dependent steering vector. Assuming the half-wavelength spacing is designed for the reference frequency $f_0$, we have
\begin{equation}
    \bs{a}^\mrm{T}(f,\phi_i) = \left[1,\;e^{\mrm{j} \pi \sin(\phi_i) \frac{f}{f_0}},\; \dots , \; e^{\mrm{j} \pi (M-1) \sin(\phi_i) \frac{f}{f_0}}\right].
\end{equation}
We further assume $\beta_i \sim \mathcal{CN}(0,1)$, $\tau_i\sim \mrm{Unif}(10^{-9}s,10^{-7}s)$, and $\phi_i\sim \mrm{Unif}(-\pi,\pi)$. 

In this work, we focus our numerical study on two frequency bands, namely a $100$ MHz frequency band around $f_0=30$ GHz, and a $50$ MHz band around $f_0=3$ GHz. These allow showcasing the main implications of the proposed analytical framework in the sub-6GHz and \ac{mmWave} bands widely employed in current wireless systems. However, the behavior of the \ac{MiLAC} processing shown in the plots is expected to generalize to other frequency bands. Note that the main effect of increasing/decreasing the center frequency is the greater/lower available bandwidth and lower/higher inductance values due to the self-resonance frequency requirements \cite{induc}, leading to approximately constant feasibility regions in \eqref{eq:ser_B_reg} and \eqref{eq:par_B_reg}. 

\subsection{Frequency-Flat Scenario}
We begin by considering an approximately frequency-flat scenario associated to a $100$ MHz band around $f_0=30$ GHz. At these \ac{mmWave} frequencies the wireless channel is typically dominated by a strong \ac{LoS} component, since the multipath components suffer from great attenuation \cite{LOS_mmWave}. We thus assume a single-tap channel by selecting $N_\mrm{MP}=1$ in \eqref{eq:sal_val_ch}. Due to the lack of delay spread, this channel will have an approximately frequency-flat response, where the only frequency dependency will come from the negligible changes in antenna spacings when measured in fractions of the wavelength. Regarding the \ac{MiLAC} processing, we assume that it has been selected to attain optimum processing at $f_0$, as discussed in Section~\ref{sec:DoFs}, and we will analyze implementations based of series, parallel, and mixed resonators as described in Section~\ref{sec:milac_imp}. For the series branches, we consider $L_\mrm{ser}=10$ nH, and, for the parallel branches, $L_\mrm{par}=0.1$ nH, which are reasonable inductance values at \ac{mmWave} frequencies. Moreover, we assume $C_{\min}=0$ and $C_{\max}=\infty$, where the extremes correspond to open and short circuits, respectively, and a characteristic impedance of $Z_0=50 \Omega$.

\begin{figure*}
\begin{subfigure}{\columnwidth}
\centering
    \includegraphics[width=\linewidth]{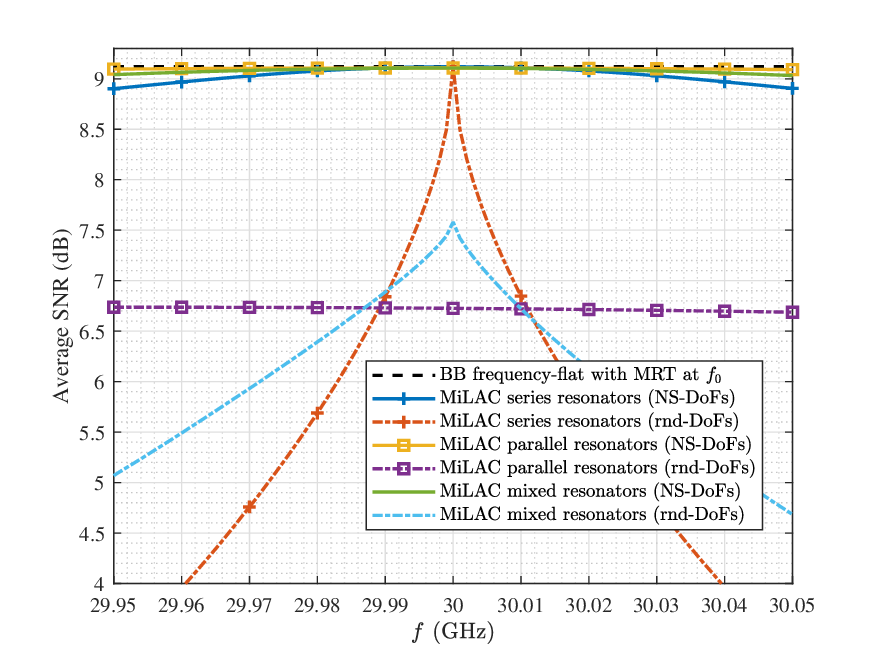}
    \caption{$M=8$}
    \label{fig:freq_flat_M8}
  \end{subfigure}
  \begin{subfigure}{\columnwidth}
  \centering
    \includegraphics[width=\linewidth]{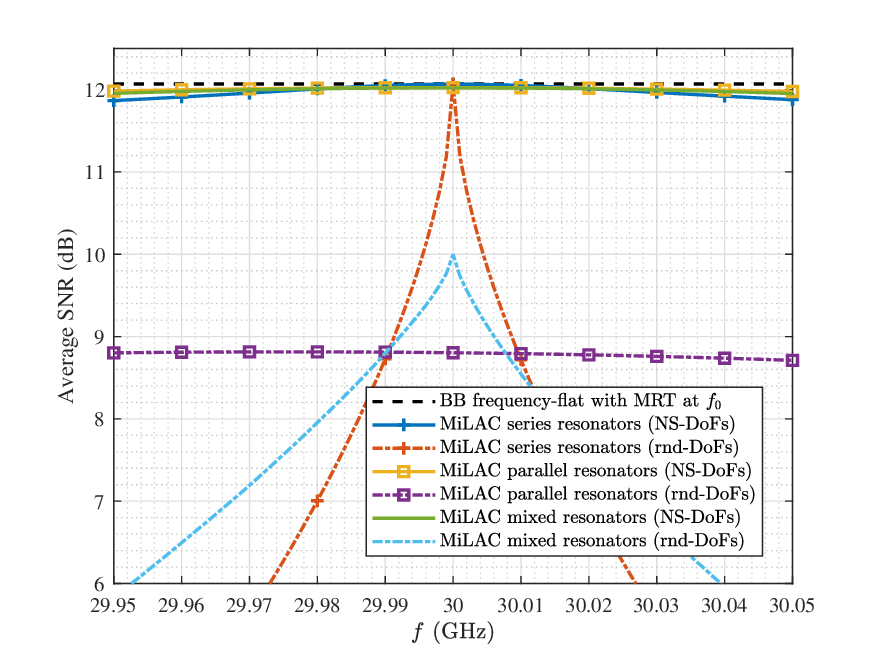}
    \caption{$M=16$}
    \label{fig:freq_flat_M16}
\end{subfigure}
 \caption{Average \ac{SNR} versus frequency for frequency-flat scenario.}\label{fig:freq_flat}
  \end{figure*}

Fig.~\ref{fig:freq_flat} shows the \ac{SNR} as a function of frequency for different \ac{MiLAC} implementations configured to attain optimal processing at $f_0$, where the \ac{DoFs} have been selected to avoid the singular points of the Cayley transform as outlined in Section~\ref{sec:freq_flat}, denoted NS-DoFs. We have fixed the input \ac{SNR} as $\frac{E_s(f)}{N_0}=1$, $\forall f$, since this corresponds to a scaling independent of the \ac{MiLAC} processing. As baseline comparison, we have considered a \ac{MiLAC}-based precoding configured to attain optimum processing at $f_0$, but where the \ac{DoFs} are chosen randomly in $\mathcal{S}_{f_0}$, denoted rnd-DoFs. We have also considered a digital approach performing frequency-flat \ac{MRT} precoding based on the channel at $f_0$. The results have been averaged across $10^4$ channel realizations. The \ac{BB} precoding curve in Fig.~\ref{fig:freq_flat} remarks the frequency-flat response of the single-tap channel. The \ac{MiLAC} with random selection of \ac{DoFs}, however, still leads to a narrowband response in the case of series resonators. When using parallel resonators with random selection of \ac{DoFs}, the response is more frequency-flat, but the lower feasibility range associated to \eqref{eq:par_B_reg} leads to sub-optimal processing at $f_0$. Our proposed configuration of the \ac{DoFs} mitigates the previous issues. Specifically, it attains a quasi-frequency-flat behavior even with series resonators, while it allows the parallel resonators to still achieve optimal processing at $f_0$ on top of maintaining the frequency-flat characteristics of parallel resonance. The \ac{MiLAC} implementations based on mixed resonators offer a trade-off between the frequency-flat behavior of parallel resonators and the reconfigurability of series resonators, especially seen in the case of random selection of \ac{DoFs}. When comparing Figs. \ref{fig:freq_flat_M8} and \ref{fig:freq_flat_M16}, we observe that the number of antennas does not have a major impact on the frequency behavior.

\subsection{Frequency-Selective Scenario}
To assess the methods proposed in Section~\ref{sec:multi_frec} exploiting the \ac{DoFs} for optimizing multi-frequency processing, we next consider a frequency selective scenario associated to a $50$ MHz band around $f_0=3$ GHz. This band is associated to multipath propagation due to the lower penetration loss and richness of scattering \cite{sub6GHz_mp}. We thus assume a large number of paths by selecting $N_\mrm{MP}=10^3$ in \eqref{eq:sal_val_ch}. This will lead to a strong frequency selectivity of the channel due to the combination of signals associated to different delays \cite{molisch}. Regarding the \ac{MiLAC} implementation, we will now consider $L_\mrm{ser}=100$ nH and $L_\mrm{par}=1$ nH for the series and parallel branches, leading to the same susceptance ranges, while the rest of considerations are maintained with respect to the frequency-flat case.

Let us now consider as performance metric the normalized \ac{SNR}, defined for each frequency $f$ as the ratio between the frequency-dependent \ac{SNR} from \eqref{eq:snr} and the optimal \ac{SNR} for each frequency, given by
\begin{equation}
    \mrm{SNR}_{\max}(f) = \frac{E_{s}(f)\Vert \bs{h}(f) \Vert^2}{N_0}. 
\end{equation}
This definition of normalized \ac{SNR} allows us to focus on the optimality of the processing at a given frequency by ignoring the variations of channel power across frequencies. Note that a normalized \ac{SNR} of 0 dB is associated to the availability of optimum processing at said frequency.

\begin{figure}
\centering
    \includegraphics[width=\linewidth]{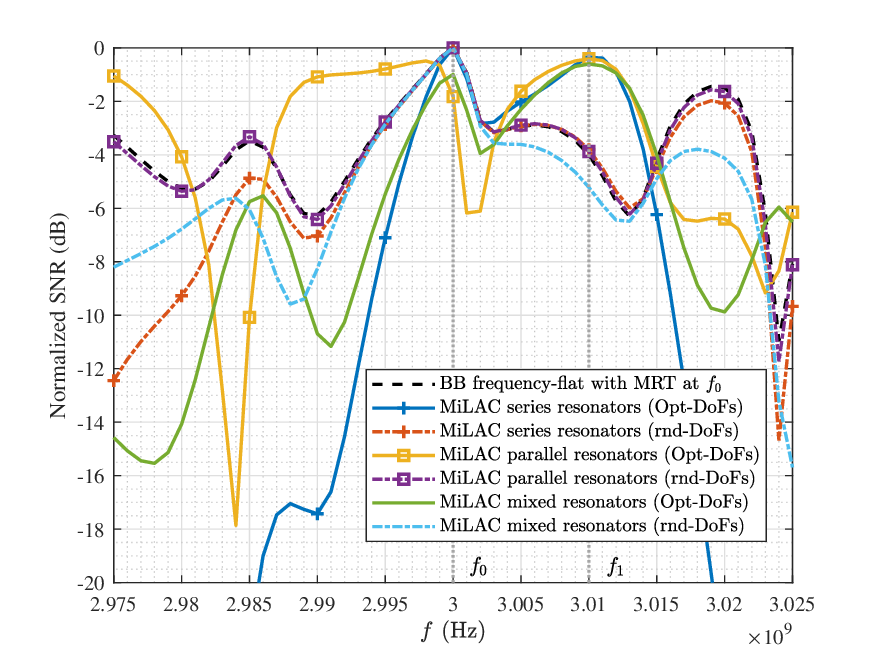}
 \caption{Normalized \ac{SNR} versus frequency for sample frequency-selective scenario with $M=2$ using exhaustive search optimization.}\label{fig:freq_sel_ex}
\end{figure}

Fig.~\ref{fig:freq_sel_ex} shows the normalized \ac{SNR} throughout the $50$ MHz band for a channel snapshot where the \ac{MiLAC} \ac{DoFs} has been optimized using exhaustive search to solve \eqref{eq:opt_prob} with $f_1=3.01$ GHz and $M=2$. In this case, the parametrization of $\mathcal{U}_{\mrm{Sym}}(M-1)$ from Proposition~\ref{prop:Unit_Sym_dep} degenerates to a single phase variable $v\in (-\pi,\pi]$ similar to that of $\varphi_0$, where we have considered an exhaustive search resolution of $0.01$ radians. We have included the same baseline comparisons as before, namely random \ac{DoFs} selection and frequency-flat \ac{BB} precoding. As we can see, the \ac{MiLAC} implementation based on series resonators attains near-optimal processing at $f_0$ and $f_1$, hinting that in some cases $\mathcal{C}^\mrm{Opt}$ in \eqref{eq:mult_freq_set} may be non-empty. For the case of parallel and mixed resonators, the performance at $f_0$ is sacrificed when solving \eqref{eq:opt_prob}, showing that $\mathcal{C}^\mrm{Opt}$ is empty in this case. This result remarks that the limiting factor is associated to the restrictions in the minimum inductance values, which limit the susceptance range \eqref{eq:par_B_reg}, since otherwise we would still be able to attain optimum processing at $f_0$. This issue would thus be solved by considering higher inductance values in parallel branches, e.g., by series connecting several inductors, or by incorporating inductance reconfigurability also such that \eqref{eq:par_B_reg} can cover approximately the whole real line. As for the baselines, the \ac{BB} processing showcases the frequency selectivity of the channel, which is closely followed by the random \ac{DoFs} \ac{MiLAC} based on parallel resonators due to its quasi-frequency-flat characteristic. In all cases, the \acp{MiLAC} with optimized \ac{DoFs} attain a significantly higher normalized \ac{SNR} at $f_1$ than the baselines.

\begin{figure}
\centering
    \includegraphics[width=\linewidth]{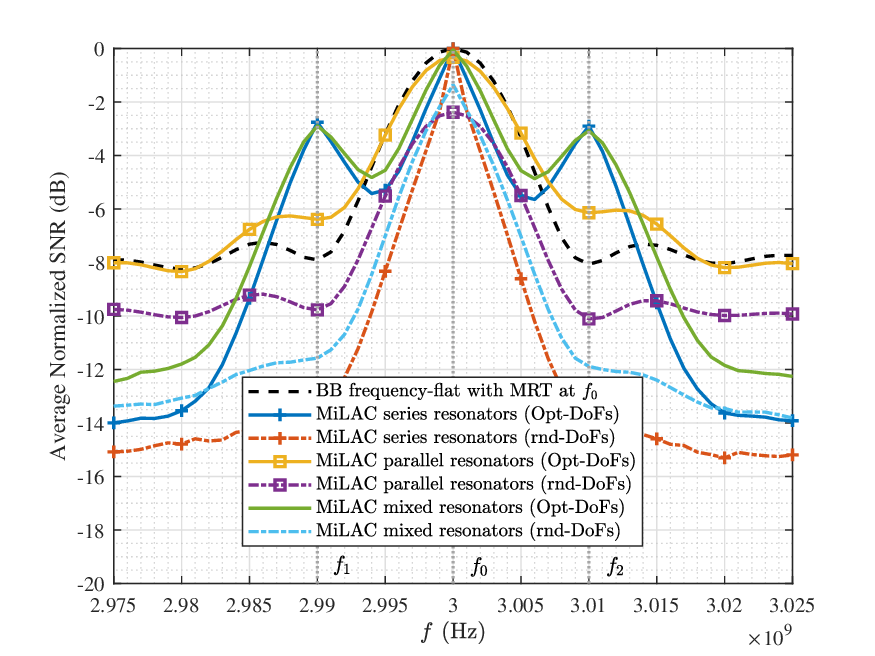}
 \caption{Average normalized \ac{SNR} versus frequency for frequency-selective scenario with $M=8$ using numerical optimization.}\label{fig:freq_sel_opt}
\end{figure}

Fig.~\ref{fig:freq_sel_opt} plots the normalized \ac{SNR} as a function of frequency where the \ac{MiLAC} \ac{DoFs} have been numerically optimized according to \eqref{eq:opt_prob_unc} for two additional frequencies, $f_1=2.99$ GHz and $f_2=3.01$ GHz with $M=8$. The optimization has been performed using standard numerical methods, namely the "fminunc" routine from MATLAB, where Proposition~\ref{prop:Unit_Sym_dep} and Corollary~\ref{cor:unc_param} have been considered for the parametrization of the \ac{DoFs}. The results have been averaged across $10^3$ channel realizations. The same baseline comparisons as before have been included. We can see again the stronger reconfigurability potential of the series resonator \ac{MiLAC} implementation, which allows for a higher \ac{SNR} at $f_1$ and $f_2$ without compromising the feasibility of the optimal processing at $f_0$. In any case, the numerical optimization is not able to attain simultaneously optimum processing at the 3 objective frequencies, leaving some ground for a further study of \eqref{eq:mult_freq_set} in future work. The optimized \ac{MiLAC} based on parallel resonators attains a significant gain with respect to that with randomly selected \ac{DoFs}, but only a small gain with respect to frequency-flat \ac{BB} precoding, showcasing its limitations in terms of adapting the processing across frequencies. On the other hand, the optimized \ac{MiLAC} based on mixed resonators is able to perform almost as good as the one based on series resonators at the objective frequencies, while having a lower processing selectivity outside these points. In all cases, the derived \ac{DoFs} are shown to be suitably exploited to improve the performance at the objective frequencies with respect to the baseline approaches.

\begin{figure}
\centering
    \includegraphics[width=\linewidth]{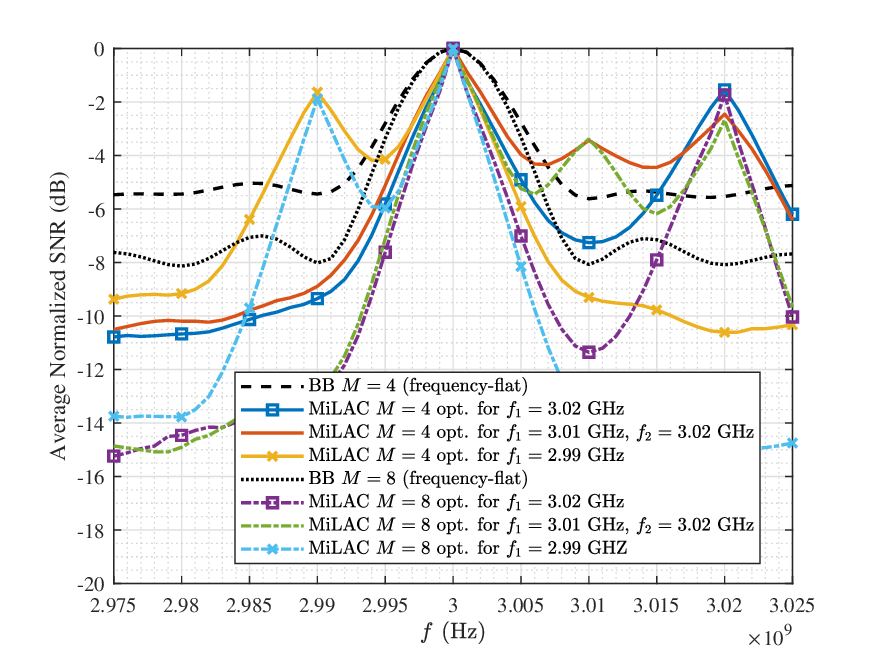}
 \caption{Average normalized \ac{SNR} versus frequency for frequency-selective scenario using numerical optimization with different antenna configuration and frequency objectives, all considering $f_0=3$ GHz and series resonators.}\label{fig:freq_sel_ser}
\end{figure}

In Fig.~\ref{fig:freq_sel_ser} we showcase the reconfiguration potential of \acp{MiLAC} based on series resonators by considering different objective frequencies and number of antennas. The same considerations as in Fig.~\ref{fig:freq_sel_opt} apply, but we have now focused exclusively in the numerically optimized \acp{MiLAC} based on series resonators and the \ac{BB} baseline. We can see that placing objective frequencies closer to $f_0$, e.g., at $f_1=2.99$ GHz, leads to similar performance as placing them slightly further, e.g., at $f_1 = 3.02$ GHz, as long as the total number of objective frequencies is maintained. However, when placing more objective frequencies, e.g., $f_1=3.01$ GHz and $f_2=3.02$ GHz, the frequencies closer to $f_0$ suffer from greater performance loss. Increasing the number of antennas has negligible effect on the normalized \ac{SNR} at the objective frequencies, while the average frequency selectivity after processing increases. Note that, since we are normalizing with respect to the optimum processing at $f_0$, the array gain is disregarded, while the performance gap between this optimum processing and the processing attained at arbitrary frequencies increases with the number of antennas. On the other hand, the additional \ac{DoFs} associated with having greater number of antennas seems to be traded-off with the difficulty to compensate for these larger gaps, as well as with the increased difficulty for numerical optimization methods to find good configurations.

\section{Conclusions}\label{sec:conc}
We have studied \ac{MiLAC}-based precoding in wideband \ac{MISO} systems using a single \ac{RF} chain. We have considered practical \ac{MiLAC} implementations based on LC resonator branches with reconfigurable capacitance. We have mapped the space of optimal \ac{MISO} baseband precoders with the space of reconfigurable capacitors, deriving the remaining \ac{DoFs} after fixing this processing at a reference frequency $f_0$. We have shown that these degrees of freedom can be used to reduce the frequency-sesitivity of \ac{MiLAC}-based processing,  leading to quasi-frequency-flat processing. We have also shown that the available \ac{DoFs} can instead be used to perform multi-frequency processing, thereby adapting the \ac{MISO} precoding at different frequencies to the frequency-selective channel. Numerical results confirm the potential of the proposed theory in both frequency-flat and frequency selective scenarios. In the frequency-flat case, the proposed closed-form \ac{DoFs} configuration attains wideband analog processing performance closely matching that of digital processing. In the frequency-selective case, we can confirm that a fully-connected passive \ac{MiLAC} can use the derived \ac{DoFs} to adapt its processing to the channel at different frequencies, offering performance gains with respect to fully digital architectures that reuse the narrowband processing throughout a wide frequency band.

The main value of the presented work lies in identifying the possibility to adapt passive analog processing to wideband scenarios by exploiting available \ac{DoFs} in fully-connected \ac{MiLAC}-inspired passive impedance networks. However, coming up with suitable optimization algorithms that can make more effective use of the derived \ac{DoFs} for improved wideband performance, as well as for alternative goals, constitutes an important direction of future work. Moreover, characterizing the \ac{DoFs} in more general \ac{MIMO} settings, where the number of input \ac{RF}-chains may be greater than 1 poses also an interesting extension. On the other hand, the theory derived in this work may be adapted to apply it in more general scenarios, e.g., when optimizing the performance in arbitrary analog computations. Such use cases may deserve a separate study in future work. Finally, the feasibility of multi-frequency optimal processing, governed by the availability of non-empty $\mathcal{C}^{\mrm{Opt}}$ in \eqref{eq:mult_freq_set}, further deserves a focused study in future work.

\appendix[Proof of Proposition~\ref{prop:DoFs_S22}]
If we substitute $\bs{u}_M=\frac{\bs{h}(f_0)}{\Vert\bs{h}(f_0)\Vert}$ in \eqref{eq:S22_comp_h} we can write
\begin{equation}\label{eq:S22_s21_unit}
\bs{S}_{22,f_0}^\mrm{H} \bs{S}_{22,f_0} + \bs{u}_{M}\bs{u}_{M}^\mrm{H}= \bs{U}\bs{U}^\mrm{H}    
\end{equation}
where we have defined the unitary matrix $\bs{U}=[\widetilde{\bs{U}}, \bs{u}_M]\in \mathcal{U}(M)$. Note that the columns of $\widetilde{\bs{U}}$ span the null-space of $\bs{h}(f_0)$. We may then identify
\begin{equation}\label{eq:S11HS11}
\bs{S}_{22,f_0}^\mrm{H} \bs{S}_{22,f_0} = \widetilde{\bs{U}}\widetilde{\bs{U}}^\mrm{H}.
\end{equation}
Given the symmetry constraint on $\bs{S}_{22,f_0}$ from  \eqref{eq:sym_S11}, we can define its Takagi factorization as
\begin{equation}\label{eq:S11}
\bs{S}_{22,f_0}=\bs{U}_{22}\bs{\Sigma}_{22}\bs{U}_{22}^\mrm{T},
\end{equation}
where $\bs{U}_{22}\in\mathcal{U}(M)$, and $\bs{\Sigma}_{22}$ is a diagonal matrix with the real non-negative singular values organized in non-ascending order throughout the diagonal. Substituting in \eqref{eq:S11HS11} yields
\begin{equation}\label{eq:u22_ut}
\bs{U}_{22}^{*}\bs{\Lambda}_{22}\bs{U}_{22}^\mrm{T}=\widetilde{\bs{U}}\widetilde{\bs{U}}^\mrm{H},
\end{equation}
where $\bs{\Lambda}_{22}=\bs{\Sigma}_{22}^2$ is the diagonal matrix of squared singular values ordered in non-ascending order. We can then identify the left-hand side with an eigenvalue decomposition of the right-hand side. In this case, since the right hand side of \eqref{eq:u22_ut} corresponds to a product of an $M\times (M-1)$ semi-unitary matrix with its conjugate transpose, the resulting matrix has its $(M-1)$ non-zero eigenvalues equal to 1, such that $\bs{\Lambda}_{22}=\diag(1,\dots,1,0)$. We may thus write
\begin{equation}
\widetilde{\bs{U}}_{22}^*\widetilde{\bs{U}}_{22}^\mrm{T}=\widetilde{\bs{U}}\widetilde{\bs{U}}^\mrm{H},
\end{equation}
where $\widetilde{\bs{U}}_{22}$ is an $M\times (M-1)$ semi-unitary matrix corresponding to the first $(M-1)$ columns of $\bs{U}_{22}$.
This leads to the restriction
\begin{equation}\label{eq:unit_DoFs}
    \widetilde{\bs{U}}_{22}=\widetilde{\bs{U}}^* \bs{Q},
\end{equation}
with $\bs{Q}\in\mathcal{U}(M-1)$ corresponding to an arbitrary unitary matrix capturing the available \ac{DoFs} in the selection of $\widetilde{\bs{U}}_{22}$. We may note that the \ac{DoFs} associated to the selection of a basis for the null-space defined by the columns of $\widetilde{\bs{U}}$ can also be parameterized by an $(M-1)\times(M-1)$ right unitary transformation. However, since unitary matrices are closed under multiplication, the resulting \ac{DoFs} can be absorbed without loss of generality by $\bs{Q}$ in \eqref{eq:unit_DoFs}.

Although $\bs{Q}$ in \eqref{eq:unit_DoFs} allows parameterizing the available \ac{DoFs} in the selection of $\bs{S}_{22,f_0}$, some of these \ac{DoFs} collapse after substitution into \eqref{eq:S11}. Specifically, considering that $\bs{\Sigma}_{22}=\bs{\Lambda}_{22}^{1/2}=\diag(1,\dots,1,0)$ since singular values are real non-negative,  we have that
\begin{equation}
\begin{aligned}
\bs{S}_{22,f_0}&=\widetilde{\bs{U}}_{22}\widetilde{\bs{U}}_{22}^\mrm{T}\\ &=\widetilde{\bs{U}}^*\bs{Q}\bs{Q}^\mrm{T}\widetilde{\bs{U}}^\mrm{H}.
\end{aligned}
\end{equation}
The \ac{DoFs} in $\bs{S}_{22,f_0}$ thus reside in the product $\bs{Q}\bs{Q}^\mrm{T}$, which can be identified with the space of unitary symmetric matrices $\mathcal{U}_\mrm{Sym}(M-1)$. Note that any $\bs{Q}_\mrm{S}\in\mathcal{U}_\mrm{Sym}(M-1)$ can be expressed as $\bs{Q}_\mrm{S}=\bs{Q}\bs{Q}^\mrm{T}$ for some $\bs{Q}\in\mathcal{U}(M-1)$, corresponding to its Takagi factorization. This shows that \eqref{eq:S22_param} provides a parametrization for $\bs{S}_{22,f_0}\in \mathcal{S}_{22,f_0}$. Furthermore, this parametrization corresponds to a bijection between $\mathcal{U}_\mrm{Sym}(M-1)$ and $\mathcal{S}_{22,f_0}$, with its inverse given by $\bs{Q}_\mrm{S}=\widetilde{\bs{U}}^T \bs{S}_{22,f_0}\widetilde{\bs{U}}$. Since both the parametrization and its inverse are restrictions of linear maps, they are smooth, and thus provide a diffeomorphism between $\mathcal{U}_\mrm{Sym}(M-1)$ and $\mathcal{S}_{22,f_0}$. This proves the diffeomorphic relation in \eqref{eq:S22_diff_Qs}, such that the total number of \ac{DoFs} in the selection of $\bs{S}_{22,f_0}\in \mathcal{S}_{22,f_0}$ coincides with the \ac{DoFs} in the selection of $\bs{Q}_\mrm{S}\in\mathcal{U}_\mrm{Sym}(M-1)$.

Let us now characterize the actual number of \ac{DoFs} available in $\mathcal{U}_\mrm{Sym}(M-1)$. This space can be associated with the orbit of $\mathbf{I}_{M-1}$ under the action 
\begin{equation}
    \alpha:\mathcal{U}(M-1)\times \mrm{Sym}_{\mathbb{C}}(M-1) \to \mrm{Sym}_{\mathbb{C}}(M-1),
\end{equation}
 defined by $\alpha(\bs{Q},\bs{A})=\bs{Q}\bs{A}\bs{Q}^\mrm{T}$. The stabilizer of $\mathbf{I}_{M-1}\in \mrm{Sym}_{\mathbb{C}}(M-1)$ under $\alpha(\cdot)$ corresponds to the subgroup of $\mathcal{U}(M-1)$ where $\alpha(\bs{Q},\mathbf{I}_{M-1})=\mathbf{I}_{M-1}$, i.e., such that  $\bs{Q}\bs{Q}^\mrm{T}=\mathbf{I}_{M-1}$, which coincides with the space of (real) orthogonal matrices $\mathcal{O}(M-1)\subset \mathcal{U}(M-1)$.\footnote{Note that the restriction $\bs{Q}^\mrm{H}=\bs{Q}^\mrm{T}$ forces all entries to be real.} By the orbit stabilizer theorem applied to the smooth manifold setting (see, e.g., \cite[Theorems~3.58, 3.62]{orb_stab_th}), we can then conclude that the space $\mathcal{U}_\mrm{Sym}(M-1)$ is diffeomorphic to the quotient space $\mathcal{U}(M-1)/\mathcal{O}(M-1)$, i.e.,
\begin{equation}
\mathcal{U}_{\mrm{Sym}}(M-1)\cong \mathcal{U}(M-1)/\mathcal{O}(M-1)    
\end{equation}
 so that the \ac{DoFs} are given by
\begin{equation}
    \dim (\mathcal{U}(M-1)/\mathcal{O}(M-1))
    =\frac{M(M-1)}{2},
\end{equation}
where we have used the property $\dim (\mathcal{X}/\mathcal{Y})=\dim (\mathcal{X})-\dim (\mathcal{Y})$.
\renewcommand{\baselinestretch}{.95}
\bibliographystyle{IEEEtran}
\bibliography{IEEEabrv,bibliography}

%

\end{document}